\documentclass[]{emulateapj}
\usepackage{natbib}
\usepackage{epsfig}
\usepackage{graphicx}
\usepackage{epstopdf}

\newcommand\icarus{{Icarus}}

\newcommand{\Npl}{N_{\rm{pl}}}
\newcommand{\mpl}{m_{\rm{pl}}}

\newcommand{\kepler}{{\it Kepler}}
\newcommand{\yr}{\rm{yr}}
\newcommand{\mcoll}{m_{\rm{coll}}}
\newcommand{\au}{\rm{AU}}
\newcommand{\stageone}{{\tt Stage\ 1}}
\newcommand{\stagetwo}{{\tt Stage\ 2}}
\newcommand{\stagethree}{{\tt Stage\ 3}}
\newcommand{\msun}{M_\odot}

\newcommand{\Rearth}{R_\oplus}
\newcommand{\Rneptune}{R_{\rm{Neptune}}}
\newcommand{\gcc}{\rm{g\:cm^{-3}}}
\newcommand{\mint}{m_{\rm{int}}}
\shorttitle{Planet-Planetesimal Scattering and MMR}
\shortauthors{Chatterjee \& Ford}

\begin{document}

\title{Planetesimal Interactions Can Explain the Mysterious Period Ratios 
of Small Near-Resonant Planets}

%\centerline{DRAFT: \today}
\author{Sourav Chatterjee}
\affil{Center for Interdisciplinary Exploration \& Research in Astrophysics (CIERA)\\Physics \& Astronomy, Northwestern University, 
Evanston, IL 60202, USA\\sourav.chatterjee@northwestern.edu}
\affil{Department of Astronomy, University of Florida, Gainesville, FL 32611, USA.}
\author{Eric B. Ford}
\affil{Department of Astronomy \& Astrophysics\\The Pennsylvania State University, 525 Davey Laboratory, University Park, PA 16802, USA}
\affil{Center for Exoplanets and Habitable Worlds\\The Pennsylvania State University, 525 Davey Laboratory, University Park, PA, 16802, USA\\eford@psu.edu}

%%%%%%%%%%%%%%%%%%%%%%%%%%%%%%%%%%%%%%%%%%%%%%%%
%ABSTRACT
%%%%%%%%%%%%%%%%%%%%%%%%%%%%%%%%%%%%%%%%%%%%%%%%
\begin{abstract}
An intriguing trend among \kepler's multi-planet systems is an overabundance 
of planet pairs with period ratios just wide of a mean motion resonance (MMR) 
and a dearth of systems just narrow of them. Traditional planet formation models 
are at odds with these observations. They are also in contrast with the period ratios 
of radial-velocity-discovered multi-planet systems which tend to 
pile up at 
$2\colon1$ MMR. We propose that gas-disk 
migration traps planets in a MMR. After gas dispersal, orbits of these trapped 
planets are altered through interaction with a residual planetesimal disk. We study 
the effects of planetesimal disk interactions on planet pairs trapped in $2\colon1$ MMR 
using planets of mass typical of the \kepler\ planet candidates (KPC) and 
explore large ranges for the mass, and density profile of the planetesimal disk. 
We find that planet-planetesimal disk interactions naturally create the observed 
asymmetry in period-ratio distribution for large ranges of planetesimal disk and 
planet properties. If the planetesimal disk mass is above a threshold of 
$\approx0.2\times$ the 
planet mass, these interactions typically disrupt MMR. Afterwards, the planets 
migrate in such a way that the final period-ratio is slightly higher than the integer 
ratio corresponding to the initial MMR. Below this threshold these interactions 
typically cannot disrupt the resonance and the period ratio stays close to the 
integer ratio. The threshold explains why the more massive planet pairs found 
by RV surveys are still in resonance. 
We encourage future research to explore how significantly the 
associated accretion would change the planets' atmospheric and surface properties.
\end{abstract}


%%%%%%%%%%%%%%%%%%%%%%%%%%%%%%%%%%%%%%%%%%%%%%%%
%INTRODUCTION
%%%%%%%%%%%%%%%%%%%%%%%%%%%%%%%%%%%%%%%%%%%%%%%%
\section{Introduction}
\label{S:intro}
\begin{figure}
\begin{center}
\plotone{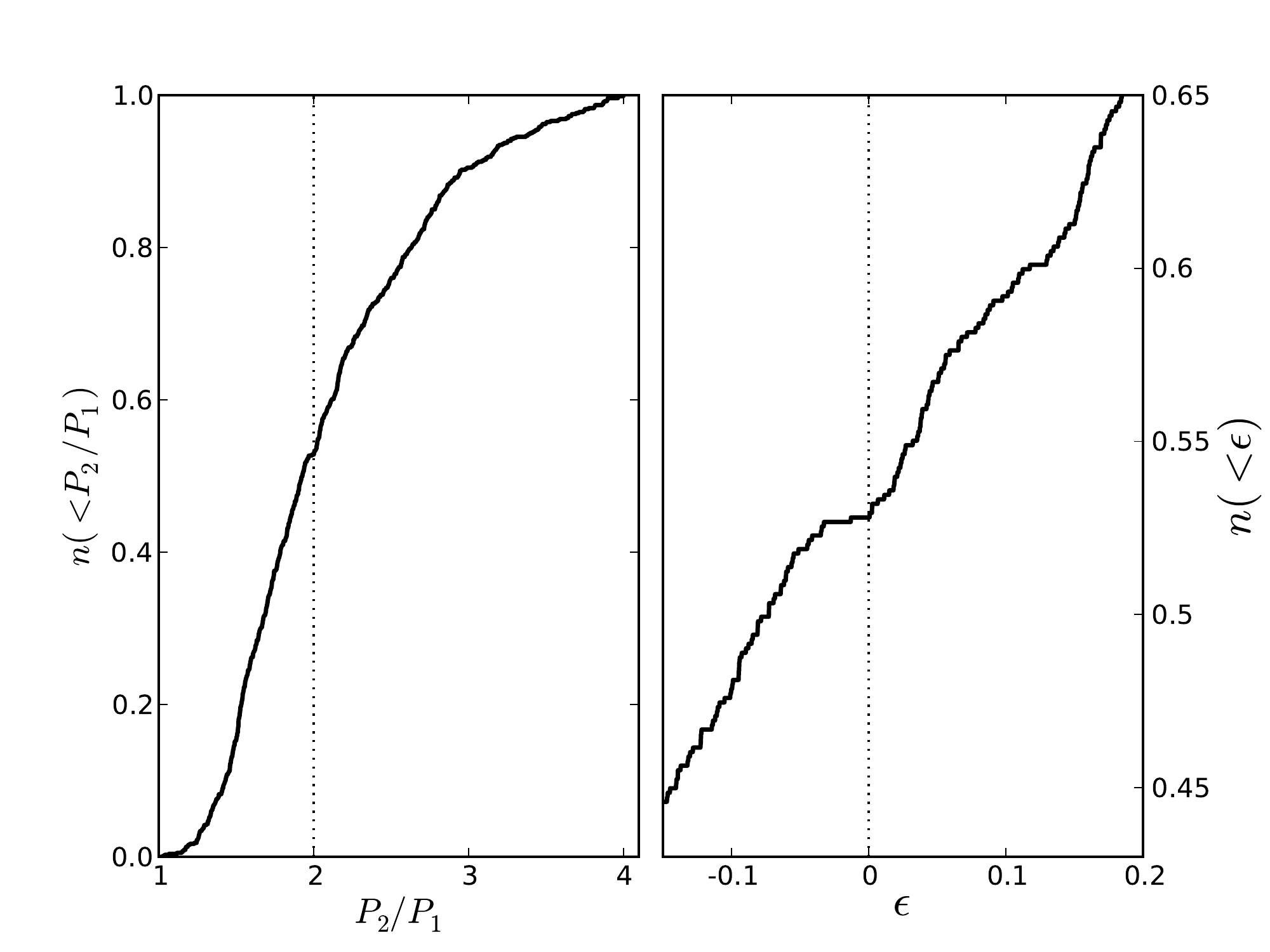}
\caption[Evolution]{
Cumulative histogram for the period ratios ($P_2/P_1$) of adjacent planet pairs discovered by \kepler\ (left). 
Right panel shows a zoomed in version of the same as a function of $\epsilon\equiv P_2/P_1-2$. The vertical 
dotted lines denote the exact positions for the $2\colon1$ MMR. 
}
\label{fig:obs}
\end{center}
\end{figure}

NASA's \kepler\ mission has revolutionized our understanding of planetary systems, 
their multiplicity, and occurrence rate 
\citep[][]{2010Sci...327..977B,2011ApJ...736...19B,2013ApJS..204...24B,2014ApJS..210...19B,2014arXiv1402.6534R,2014ApJ...790..146F}. 
Most of the candidates 
have radii between $\sim\Rearth$ to Neptune radius ($\Rneptune$). Several trends, not all 
a-priori expected, have emerged among this new class of small (presumably low-mass) planet 
population. One of the most prominent trends is the existence of extremely compact, well 
aligned, short period multi-transiting systems \citep[e.g.,][]{2012ApJ...761...92F,2012ApJ...751..158H,2013ApJ...775...53H,2013MNRAS.431.3444C,2014ApJ...780...53C}. Another trend 
is that although the spacings between planet pairs among most KPCs seems random, there is 
a clear overabundance of pairs just wide of major MMRs including $2\colon1$, and $3\colon2$, 
(Figure\ \ref{fig:obs}) and a lack of planet pairs just inside of these resonances 
\citep{2011ApJS..197....8L,2014ApJ...790..146F}. 
This feature in the period-ratio distribution in the KPCs is in drastic contrast to that observed 
in the planet populations discovered via RV 
surveys \citep{2006ApJ...646..505B} which preferentially find 
much higher mass planets, and shows 
a clear peak very near the $2\colon1$ MMR. 
Tests for the statistical significance of this apparent overabundance indicate a clear deficit of planet pairs 
interior to $2\colon1$ and an overabundance near period ratio of $2.2$ \citep{2014arXiv1409.3320S}.   

Smooth gas-disk driven 
convergent migration is expected to trap planets into MMRs resulting in period ratios 
that are very close to the exact integer ratios, and this is thought to be responsible for the giant planet 
MMR pairs \citep[e.g.,][]{2002ApJ...567..596L,2013apf..book.....A}. 
The orbital period ratios of 
a pair of planets near the $j+1\colon j$ MMR usually have small ($10^{-3}$) offsets 
from the exact integer ratios, 
$\epsilon\equiv P_2/P_1 - (j+1)/j\lesssim10^{-3}$, where 
$P_i$ denotes the orbital period 
of the $i^{th}$ planet (e.g., $\epsilon = P_2/P_1 - 2/1$ for $2\colon1$ MMR). 
Planet-planet 
scattering, on the other hand, brings large changes to 
the orbits of the planet pairs initially trapped into a MMR and likely 
dramatically increases the relative inclinations of these orbits making multi-transiting 
configuration impossible \citep[e.g.,][]{1996Sci...274..954R,2008ApJ...686..580C,2008ApJ...686..603J,2011ApJ...742...72N,2012ApJ...751..119B}. 
Hence, the process responsible 
for the observed large ($\lesssim0.2$) positive $\epsilon$ distribution 
must not be as smooth as gas-disk migration and 
also not as strong as planet-planet scattering. 
The mechanism responsible for this feature must be 
fairly common during planetary system formation, as evidenced by the 
high fraction of near-resonance KPC pairs exhibiting this feature. This process 
must also be significantly more powerful for low-mass planets compared 
to the giants where this feature has not been observed. 

Several theories have been proposed to explain this feature, 
\citep{2012ApJ...756L..11L,2012MNRAS.427L..21R,2013AJ....145....1B,2013ApJ...770...24P,2014A&A...570L...7D}. 
For example, resonance repulsion in presence of tidal damping 
\citep{2012ApJ...756L..11L,2013AJ....145....1B} creates positive 
$\epsilon$, but predicts values that are at least an order of magnitude too small suggesting 
additional dissipative processes are at play \citep{2013ApJ...774...52L}. In situ growth 
of planets via planetesimal accretion \citep{2013ApJ...770...24P} assumes idealized 
and likely unphysical initial conditions since planetesimal accretion naturally results in 
changes in the semi major axis ($a$). 
Gas disk interactions in presence of turbulence may result in positive $\epsilon$, but the 
results are strongly dependent on the strength of such turbulence which is largely uncertain 
and highly variable. 

This leads us to search for a mechanism that can easily create asymmetric shifts in periods from exact 
MMRs, that is expected to be ubiquitous within the core-accretion paradigm of planet 
formation, and that is not overly sensitive on the details of the initial conditions. 
We propose that gas-disk driven migration traps some planet pairs in MMR with 
low $\epsilon$, as expected from conventional theories \citep[e.g.,][]{1996Natur.380..606L,1980ApJ...241..425G,2000ApJ...540.1091B}. 
After gas dispersal, these planets interact with planetesimals from a residual disk, 
expected to be 
present from the core-accretion paradigm of planet formation. 
These interactions are stochastic, but of much weaker strength than planet-planet scattering. 
Planetesimal driven migration (in presence or absence of a gas disk) is a well studied 
process and has been long identified as 
an important ingredient in understanding the formation and evolution of planets, 
especially, in the context of the outer Solar system 
\citep[e.g.,][]{1984Icar...58..109F,1999AJ....117.3041H,2007prpl.conf..669L,2009Icar..199..197K,2011ApJ...735...29B,2012ApJ...758...80O,2014Icar..232..118M}. 
More recently, numerical results suggest that planetesimal 
disk scattering can significantly alter $\epsilon$ for some specific \kepler\ systems 
\citep{2013MNRAS.432.1196M}. 

We systematically study the effects of 
planetesimal disk interactions on resonant planet pairs. In particular, we focus on the 
$2\colon1$ MMR in this study since the difference in the period ratio distribution between 
low-mass planets and giant planets is the most dramatic near this MMR. 
In \S\ref{S:modeling} we 
describe our numerical setup, the explored parameter space, and explain the choices of 
our initial conditions. In \S\ref{S:results} we describe our key results. Finally, we 
discuss the strengths and weaknesses of this process and explain implications that can 
be observationally tested in \S\ref{S:discussion}.

%%%%%%%%%%%%%%%%%%%%%%%%%%%%%%%%%%%%%%%%%%%%%%%%
%THEORETICAL MODEL
%%%%%%%%%%%%%%%%%%%%%%%%%%%%%%%%%%%%%%%%%%%%%%%%
%
%%%%%%%%%%%%%%%%%%%%%%%%%%%%%%%%%%%%%%%%%%%%%%%%
%FIGURE 1
%%%%%%%%%%%%%%%%%%%%%%%%%%%%%%%%%%%%%%%%%%%%%%%%
\begin{figure*}
\begin{center}
\plotone{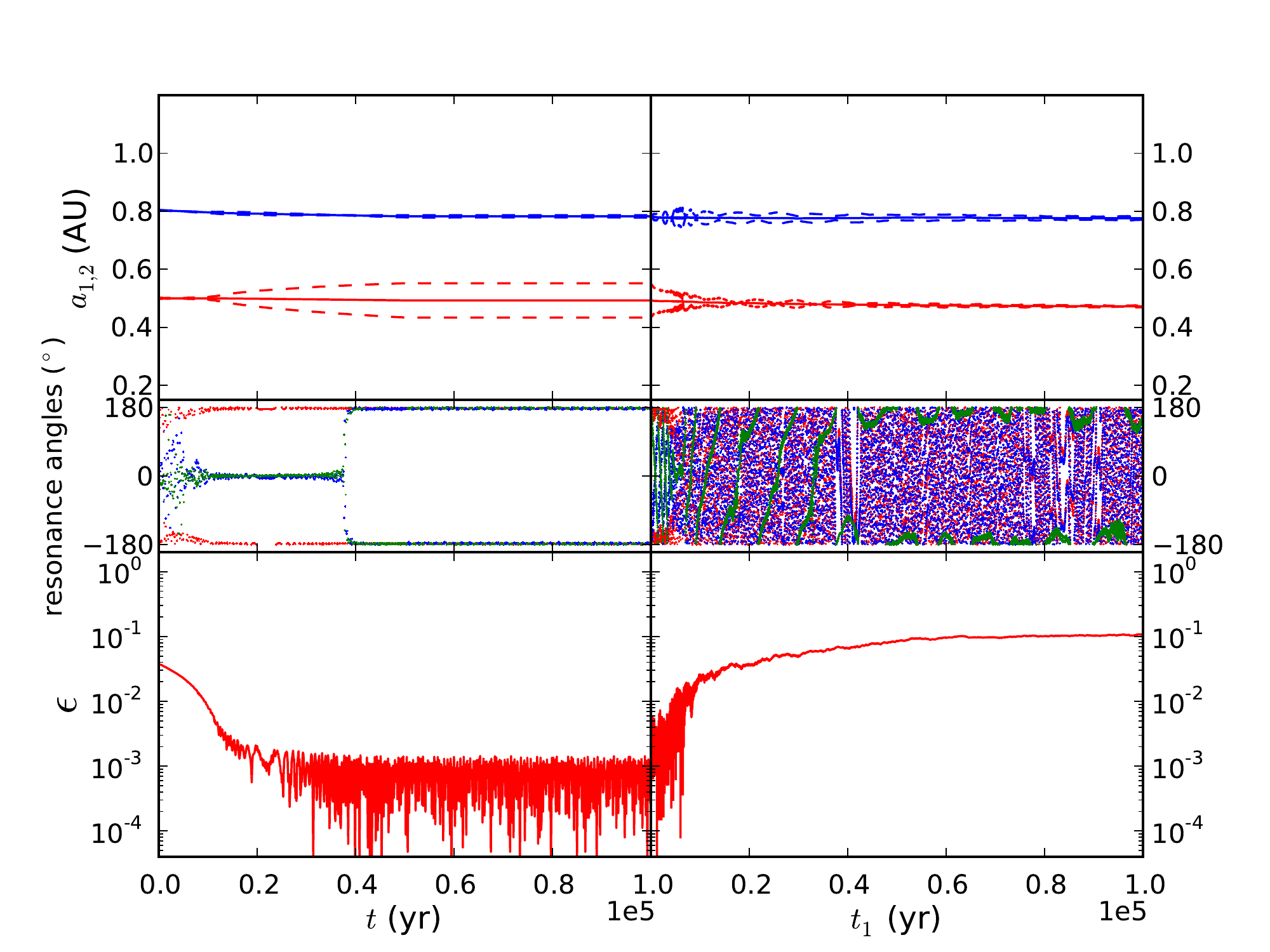}
\caption[Evolution]{
Example orbital evolution of resonant planet pairs. 
For this example we choose one realization with initial properties $m_1/m_2 = 1$, 
$m_d/m_p=0.5$, and $\alpha=-1.5$. Top panels show the evolution of the 
semimajor axes (solid), pericenter and apocenter distances (dashed). The middle 
panels show the evolution of three resonant angles for $2\colon1$ resonance. 
The bottom panels show the evolution of $\epsilon=P_2/P_1-2/1$. 
Left panels: The evolution of planet pairs during \stageone\ (\S\ref{S:stage1}). 
The resonant angles are initially circulating indicating non-resonance. These 
angles start librating when convergent migration traps the two orbits in $2\colon1$ 
MMR. Orbital eccentricities increase due to resonance trapping. The offset 
$\epsilon$ steadily decreases until it reaches $\epsilon\sim10^{-3}$. 
Right panels: The evolution of the same resonant planets in presence of a 
planetesimal disk. Planets interact with the planetesimals. Orbital eccentricities 
are damped. Initially, $\epsilon$ changes chaotically due to individual 
planet-planetesimal interactions. After several such interactions the overall 
perturbations disrupt the resonance indicated by recirculation of the resonant 
angles, in this example, at $t_1 \sim 5 \times 10^3\,\yr$. Once the resonance 
is disrupted, $\epsilon$ approaches $\approx0.1$.
}
\label{fig:evolution}
\end{center}
\end{figure*} 
%

%%%%%%%%%%%%%%%%%%%%%%%%%%%%%%%%%%%%%%%%%%%%%%%%
%NUMERICAL MODELING
%%%%%%%%%%%%%%%%%%%%%%%%%%%%%%%%%%%%%%%%%%%%%%%%
\section{Numerical Modeling}
\label{S:modeling}
A fully realistic numerical model of systems containing two planets trapped in a MMR emerging 
from a full protoplanetary disk with planets, planetesimals, and gas is impractical and beyond the 
scope of this study. 
Instead, 
we use pure $N$-body models with initial conditions generated to mimic the expected properties of 
a system emerging from a dissipative gas disk. 
In order to create reasonable initial conditions for the planet 
and planetesimal orbits we adopt a 3-stage algorithm. In \stageone\ we create orbital properties 
for planet pairs in $2\colon1$ MMR by migrating the outer planet inwards (\S\ref{S:stage1}). 
In \stagetwo\ we create initial 
planetesimal disk properties that are dynamically consistent 
with the resonant planet pair properties (\S\ref{S:stage2}). These two steps create 
initial conditions for the final and main stage of models including two resonant planets and a dynamically 
active planetesimals disk (\stagethree; \S\ref{S:stage3}). 
We test whether late stage (after gas disk is dispersed) interactions between the resonant 
planets and a residual planetesimal disk can create large positive offsets from initial integer ratio of 
periods expected of the MMR. 
We use 
the hybrid integrator of MERCURY6.2 \citep{1999MNRAS.304..793C} for all stages of 	our 
simulations. We use a planetary density of $1.64\,\rm{g\:cm^{-3}}$, typical of \kepler's 
multi-planet systems \citep[e.g.,][]{2014ApJ...787..173H,2014ApJ...783L...6W,2015ApJ...798L..32C}, 
to calculate the planet sizes for all our simulations. 
The planetesimal density is chosen to be $6\,\rm{g\:cm^{-3}}$. 
Below we describe the detailed numerical treatments for each 
stage of our simulations. 

%%%%%%%%%%%%%%%%%%%%%%%%%%%%%%%%%%%%%%%%%%%%%%%%
%STAGE 1
%%%%%%%%%%%%%%%%%%%%%%%%%%%%%%%%%%%%%%%%%%%%%%%%
%
\subsection{Stage\ 1: Trapping planet pairs into $2:1$ mean motion resonance}
\label{S:stage1}
In this stage our initial properties consist of two planets of masses $m_1$ and $m_2$ 
(indices are counted from inside out). The initial semimajor axis for the inner planet is 
$a_1 = 0.5\,\au$. Initial semimajor axis of the outer planet $a_2$ is chosen to be $0.02\,\au$ 
outside the exact period ratio of $P_2/P_1 = 2$. Initially both orbits are circular. Orbital planes of the two planets are aligned 
initially. We choose other orbital phase angles uniformly from their full ranges. The two planets 
are evolved using the hybrid integrator. We mimic gas disk damping and convergent 
migration by applying a forced migration using 
$\dot{a_2}=10^{-6}\,\au yr^{-1}$ and $\dot{e_2}=10^{-4}\,\yr^{-1}$ on the 
outer planet \citep{2002ApJ...567..596L}. The outer planet moves in smoothly and the two 
orbits get trapped into $2\colon1$ MMR at $t\approx2\times10^{4}\,\yr$. The trapped planets 
move further inwards together for another $2\times10^{-4}\,\yr$, when the $\dot{a_2}$ and 
$\dot{e_2}$ terms are switched off. We follow the evolution of the resonant planets till 
$t=10^5\,\yr$. The left panels in Figure\ \ref{fig:evolution} 
show an example of the evolution of the planetary orbital properties in \stageone. 
Initially the resonant angles circulate and the planets are not in resonance. 
Resonant angles start liberating as the planets are trapped into $2\colon1$ MMR. The 
value of $\epsilon$ steadily decreases and reaches a small value $\epsilon\lesssim10^{-3}$. 
At the end of \stageone\ we note the orbital properties of the planet pair and thus initial 
orbital properties for a resonant planet pair are created. 

%
%%%%%%%%%%%%%%%%%%%%%%%%%%%%%%%%%%%%%%%%%%%%%%%%
%FIGURE 2
%%%%%%%%%%%%%%%%%%%%%%%%%%%%%%%%%%%%%%%%%%%%%%%%
%
\begin{figure}
\begin{center}
\plotone{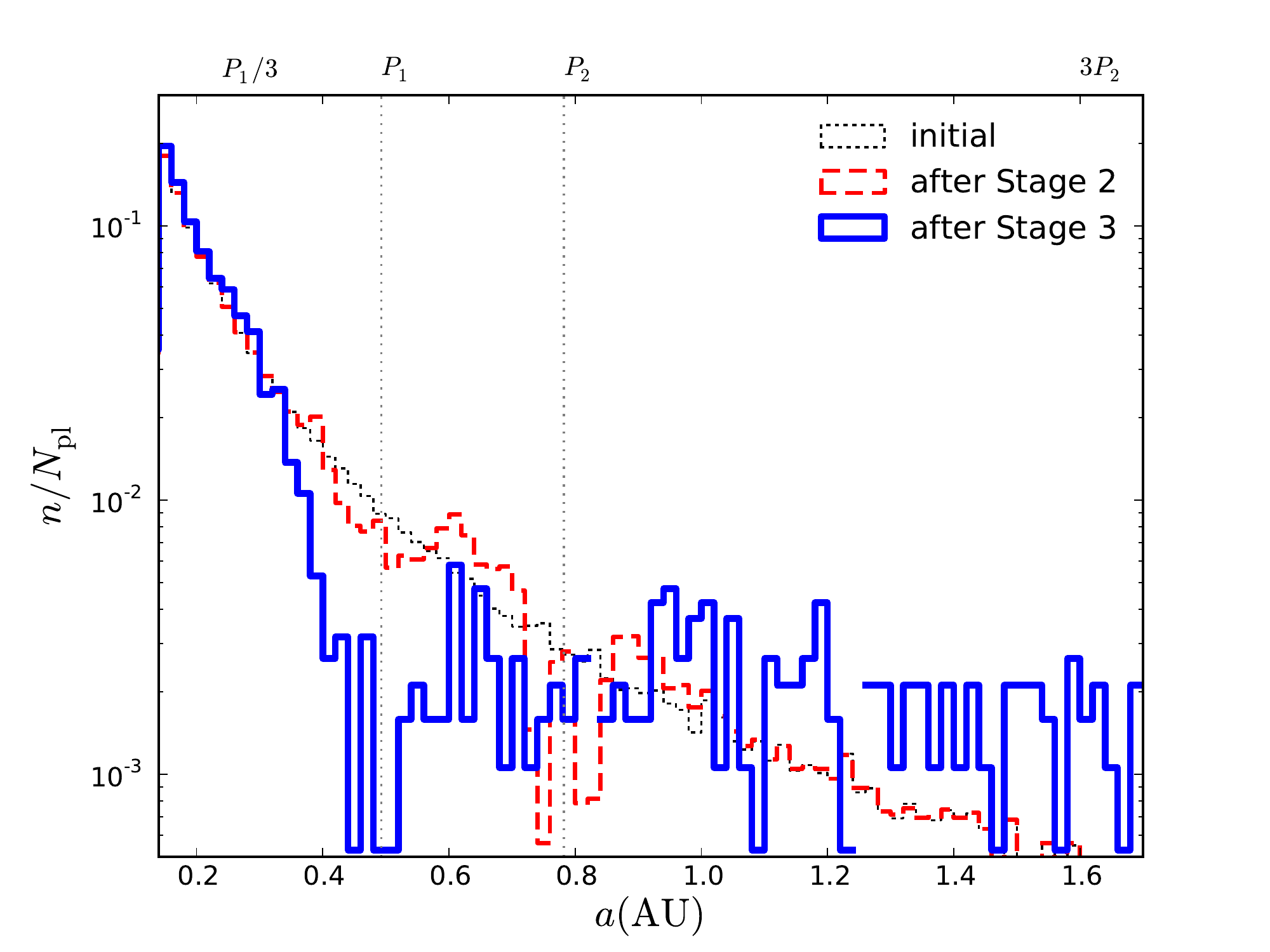}
\caption[Example disk profiles $dn/da$]{
Example evolution of the planetesimal disk profile for the same model as in 
Figure\ \ref{fig:evolution}. Dotted (black), dashed (red), and thick solid (blue) histograms 
show the initial undisturbed power-law profile, the profile for all planetesimals that 
survived until the end of \stagetwo, and the final profile for planetesimals at the end 
of \stagethree, 
respectively. The vertical dotted lines show the 
positions of the planets at the beginning of \stagethree. 
The positions of $P_1/3$ and 
$3P_2$ are shown as well, for reference since these values are used to determine the positions 
of the disk edges (\S\ref{S:stage2}).      
Each histogram is normalized by the total number of planetesimals  
at each stage ($10^5$, $99653$, and $1892$ for initial, after \stagetwo, and after \stagethree, respectively). The presence of the planet pair alter the power-law profile. So the planetesimal 
surface density exhibits 
dynamical features at the end of \stagetwo. At the end of \stagethree\ only a small number 
of planetesimals remain close to the planets. %
}
\label{fig:profile}
\end{center}
\end{figure} 
%

%
%%%%%%%%%%%%%%%%%%%%%%%%%%%%%%%%%%%%%%%%%%%%%%%%
%STAGE 2
%%%%%%%%%%%%%%%%%%%%%%%%%%%%%%%%%%%%%%%%%%%%%%%%
%
\subsection{Stage\ 2: Creation of dynamically consistent planetesimal disk} 
\label{S:stage2}
We first choose a disk described by a simple power-law of the form $d\Sigma/da=a^\alpha$. 
We ascertain that the disk edges are sufficiently far from the planets to avoid unpredictable 
edge effects by setting the inner disk edge at $0.01\,\au$ interior to the $a$ value for $1\colon3$ 
period ratio with the inner planet. Similarly, we set the outer edge of the disk at $0.01\,\au$ 
exterior to the $a$ value for the $3\colon1$ period ratio with the outer planet orbit. The planetesimal 
positions are selected consistent with the power-law profile. The planetesimal eccentricities ($e$)
are drawn uniformly in the range $0$ to $10^{-3}$. The initial orbital inclinations are drawn 
randomly to vary between $-1^\circ$ and $1^\circ$. Other orbital phase angles for the planetesimals 
are chosen randomly in their full ranges.   

In a real system, while a gas disk is present, it can damp dynamical excitations in the planetesimal 
orbits caused by the resonant planets. However, planetesimal orbits that are unstable 
on very short timescales will still be disrupted via collisions and scattering due to strong gravitational interactions 
with the planets \citep[e.g.,][]{2010ApJ...714..194M}. 
As the system emerges out of the gas disk, the planetesimal 
disk will not remain a pure power-law, but will have features such as a decrease in density 
near the planets that perturb the initial power-law, resonance features, and a larger dispersion 
in $e$.  

We model the planetary perturbations to the planetesimal disk surface density profiles 
in the following way. We collect final orbital properties of the resonant planet 
pairs created in \stageone. We evolve these planets with a swath of planetesimals 
with properties obtained using a particular value of $\alpha$ and disk edges as 
described above for $100\,\yr$. At this stage we treat all planetesimals as test particles, 
so the planets can dynamically alter the planetesimal orbits, but the planetesimals cannot 
change planet properties. After $t=100\,\yr$ we stop the integrations and collect orbital 
properties of the surviving planetesimals. We create a large database of $N\sim10^5$ 
surviving planetesimal orbits for each $\alpha$ and each combination of $m_1$ and 
$m_2$. The database will not include planetesimals on orbits that rapidly become 
unstable, for example, via physical collisions with the planets. However, if planetesimals 
can survive in specific orbital configurations, such as $1\colon1$ resonance, 
horse-shoe or tadpole orbits, then such orbits will be naturally populated among the 
orbital database of surviving planetesimals. Hence, at the end of \stagetwo, a database 
of planetesimal orbits is created that is dynamically consistent for each combination 
of resonant planet pair mass and initial planetesimal disk surface density profile 
(e.g., Figure\ \ref{fig:profile}). 

We expect that shortly before gas disk dispersal $e$-damping may not be efficient 
and eccentricities of planetesimal orbits near the planets will likely grow 
\citep{2010ApJ...714..194M}. Therefore, 
during \stagetwo\ we do not apply any forced damping. As a result,  
planetesimal eccentricities grow freely via encounters with the planets. The exact values 
depend on the details of the initial planet and planetesimal disk properties. We find that 
the median $e$ for all planetesimal orbits at the end of \stagetwo\ is 
$\sim(6.1\pm0.5)\times10^{-3}$. For planetesimal orbits within $a_1 - 10R_{H,1}$ and $a_2 + 10R_{H,2}$ 
($R_{H,i}$ is the Hill radius of the $i^{th}$ planet), a region within which almost all planetesimals 
will have a strong encounter with the planets during \stagethree,
the median $e$ is a little higher $\sim(1.5\pm0.2)\times10^{-2}$. Here the error bars denote 
the standard deviation in the $e$-distributions.  

Our simulations do not explicitly include a gas disk. Rather, they are designed to 
imitate the expected effects of a gas disk, namely, removal and alterations of only those orbits that 
are unstable on short timescales ($\sim 100\,\yr$ is our choice), and preservation of orbits that 
are not unstable on these short timescales.  
The integration stopping time of $100\,\yr$ in \stagetwo\ is somewhat arbitrary since it is 
hard to predict the stabilizing effect of the gas disk and exactly which planetesimals will 
be unstable even while the gas is present. Our choice of $100\,\yr$ results in removing 
planetesimal orbits unless they are stable for $\gtrsim3\times10^2$ orbits of the 
inner planets. Hence, further interactions at orbital timescales are not expected. Integrating 
longer would not change 
the overall qualitative results as long as there are enough 
planetesimals in the disk to interact with the planets at the end of this stage.  

%
%%%%%%%%%%%%%%%%%%%%%%%%%%%%%%%%%%%%%%%%%%%%%%%%
%STAGE 3
%%%%%%%%%%%%%%%%%%%%%%%%%%%%%%%%%%%%%%%%%%%%%%%%
%
\subsection{Stage\ 3: Evolution of initially resonant planet pairs in presence of a massive planetesimal disk} 
\label{S:stage3}
This is the main stage of our study where simulations follow the evolution of planet pairs in 
resonance embedded in a dynamically consistent planetesimal disk. 
We randomly choose 
$\Npl=2\times10^3$ planetesimal orbits from the orbital databases we create in \stagetwo\ 
for each combination of a resonant planet pair and initial power-law exponent $\alpha$. 
We extract the orbital properties for the resonant planet pair at the end of \stagetwo. 
We evolve the planets and planetesimals together. 
Now, the planetesimals are treated as pseudo-test particles, 
such that planetesimals interact with the planets, but they do not interact 
between themselves. This approximation 
reduces the computational cost and is not expected to change our results significantly. 
All planetesimals are of equal mass in each simulated model. The mass of each 
planetesimal is set to achieve the desired ratio between 
the total planetesimal disk mass ($m_d$) and the total planet mass $m_p=m_1+m_2$. 
We stop our integrations at $t_1 = 10^5\,\yr$.  We have tested that longer 
integrations and integrations with larger $\Npl$ (at fixed $m_d$) do not alter the 
results in a statistically significant way. 
At the end of the integrations we retrieve the orbital properties of the planets 
and all planetesimals. 
We study the evolution of orbital properties of the initially resonant planets including 
$\epsilon$ and the resonant angles.    

%
%%%%%%%%%%%%%%%%%%%%%%%%%%%%%%%%%%%%%%%%%%%%%%%%
%PARAMETER SPACE
%%%%%%%%%%%%%%%%%%%%%%%%%%%%%%%%%%%%%%%%%%%%%%%%
%
\subsection{Exploration of Parameter Space} 
\label{S:param_grid}
We restrict our study to the $2\colon1$ MMR since 
$\epsilon$-values for \kepler\ systems show the strongest difference 
from RV-discovered systems near period ratio of $2$, primarily due to a large number 
of RV-discovered giant planet pairs near a period ratio of $2$. 

We choose the planet masses in the following way. The more massive planet's mass is 
set to $5\times10^{-5}\ \msun \approx 1\ M_{\rm{Neptune}}$. 
The mass of the other planet, is determined using a grid of $m_1/m_2$ values. 
We vary $m_1/m_2 = 0.1, 0.2, 1, 5$, and $10$. 
For example, for $m_1/m_2 = 5$, $m_1 = 5\times10^{-5}\ \msun$, $m_2 = 10^{-5}\ \msun$ 
and for $m_1/m_2=0.2$, $m_1=10^{-5}\,\msun$, $m_2=5\times10^{-5}\,\msun$. 

We vary the disk density profile given by $d\Sigma/da \propto a^\alpha$ by choosing $\alpha$ 
in a large range between $-2.5$ to $3$ using a grid step-size of $0.5$. 
For each combination of $m_1$, $m_2$, and $\alpha$ we vary $m_d/m_p$ 
between $0.1$ to $1$ with a 
grid step-size of $0.1$. In addition, we include $m_d/m_p = 1.5$. 
To address the inherent statistical fluctuations 
we simulate $4$ realizations for each choice of $m_1/m_2$, $\alpha$, and $m_d/m_p$. Thus, 
our results are based on a large ensemble ($2620$ simulations in \stagethree, each with 
two planets trapped in $2\colon1$ MMR and $2\times10^3$ pseudo-test particles) 
of models. 

%
%%%%%%%%%%%%%%%%%%%%%%%%%%%%%%%%%%%%%%%%%%%%%%%%
%PLANETESIMAL NUMBERS
%%%%%%%%%%%%%%%%%%%%%%%%%%%%%%%%%%%%%%%%%%%%%%%%
%
\subsection{Number of planetesimals} 
\label{S:npl}
The choice of $\Npl$ in \stagethree\ is somewhat arbitrary 
and guided by the following considerations. In this study we will test 
whether the cumulative effect of many small perturbations from several 
planetesimals can eventually disrupt the $2\colon1$ resonance between a pair of 
planetary orbits, and leave them close to the initial period commensurability 
with $\epsilon \sim 0.01$ to $0.2$. Hence, it is essential to ensure that 
the perturbation caused by an individual planet-planetesimal interaction 
is sufficiently small and cannot disrupt the resonance. For a given $m_d/m_p$ 
the higher the $\Npl$, the smoother the evolution. 
However, the computational cost scales as $\sim \Npl$ for the 
pseudo test-particles. Combined with the requirement of small timesteps (we use $1\,\rm{day}$) to 
increase accuracy and a sufficiently long integration time, making $\Npl$ very large is 
computationally impractical. Fortunately, analytical considerations from resonance 
theory can guide us to determine a sufficient value of $\Npl$ while keeping the 
computational costs reasonable.  

We require that the 
maximum possible fractional change $\delta a/a$ 
to a planet's orbit caused by dynamical interaction with 
a single planetesimal is lower than the resonance width $\Delta a/a$ of the MMR. 
The libration width $|\Delta a/a|$ for $2\colon1$ MMR in the restricted three-body case is 
\begin{equation}
\left| \frac{\Delta a}{a} \right| \geq 78.43 \frac{m_p}{m_\star}
\end{equation}
where $m_p$ and $m_\star$ are the planet mass and the star mass, 
respectively \citep{1999ssd..book.....M}. 
The maximum change in $\delta a/a$ from a single planet-planetesimal 
interaction is 
\begin{equation}
\left| \frac{\delta a}{a} \right|_{\rm{pl}} \approx 2 \frac{\mpl}{m_\star} %
\end{equation}
where $\mpl$ is the mass of the planetesimal. 
Hence, our requirement is satisfied if  
\begin{equation}
\frac{\mpl}{m_p} \leq 39.22 \frac{m_p}{m_\star}.  
\end{equation}
Using our model assumptions of $m_\star=1\,\msun$, $m_p=5\times10^{-5}\,\msun$, 
and for equal mass planetesimals, the above condition is equivalent to 
$\Npl\gtrsim5\times10^2$. Numerical tests of \stagethree\ simulations using varying $\Npl$ 
verify  that any $\Npl\gtrsim 5\times 10^2$ gives statistically indistinguishable results, and 
$\Npl\gtrsim10^3$ makes the evolution fairly smooth. Hence, 
the choice of $\Npl = 2\times10^3$ in our models is adequately high for this study. 
Using $\Npl=2\times10^3$ and by treating 
them as pseudo-test particles, simulations in \stagethree\ took a total of $\sim5\times10^5$ 
CPU hours to complete.     
%

%
%%%%%%%%%%%%%%%%%%%%%%%%%%%%%%%%%%%%%%%%%%%%%%%%
%FIGURE CEWHICH
%%%%%%%%%%%%%%%%%%%%%%%%%%%%%%%%%%%%%%%%%%%%%%%%
%
\begin{figure}
\begin{center}
\plotone{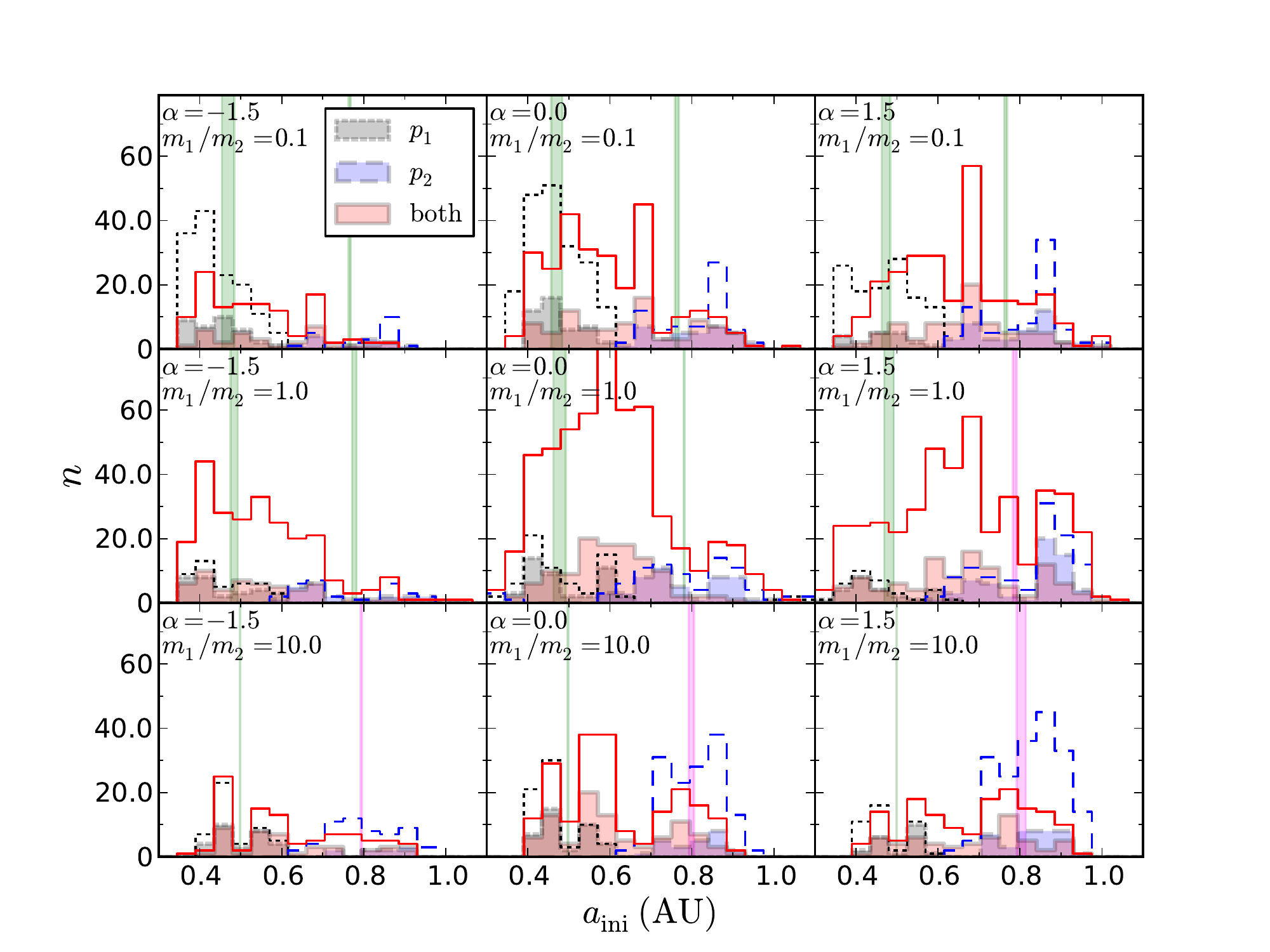}
\caption[$P_2/P_1$ for all $\alpha$]{
Distributions for the initial positions of planetesimals 
that came within $1R_H$ of only planet 1 (black dashed), only planet 2 (blue long-dashed), 
and both planets (red solid) during \stagethree\ evolution. Filled histograms denote the same, 
but includes only planetesimals that collided with one of the planets in the end. Each panel 
shows a particular combination of 
initial $m_1/m_2$ and $\alpha$ (noted in each panel). All panels are for $m_d/m_p=0.5$. The 
vertical shaded regions are bounded by the initial and final planet positions. Green and red 
for the shaded regions denote inward and outward migration, respectively. In all of these example 
cases the inner planet migrated inward. When the outer planet also migrates inward, it does so 
at a lower rate compared to the inner planet resulting in growth of $\epsilon$ in all of these cases.   
}
\label{fig:cewhich}
\end{center}
\end{figure} 
%

%
%%%%%%%%%%%%%%%%%%%%%%%%%%%%%%%%%%%%%%%%%%%%%%%%
%FIGURE PRATIO HIST ALL ALPHA
%%%%%%%%%%%%%%%%%%%%%%%%%%%%%%%%%%%%%%%%%%%%%%%%
%
\begin{figure*}
\begin{center}
\plotone{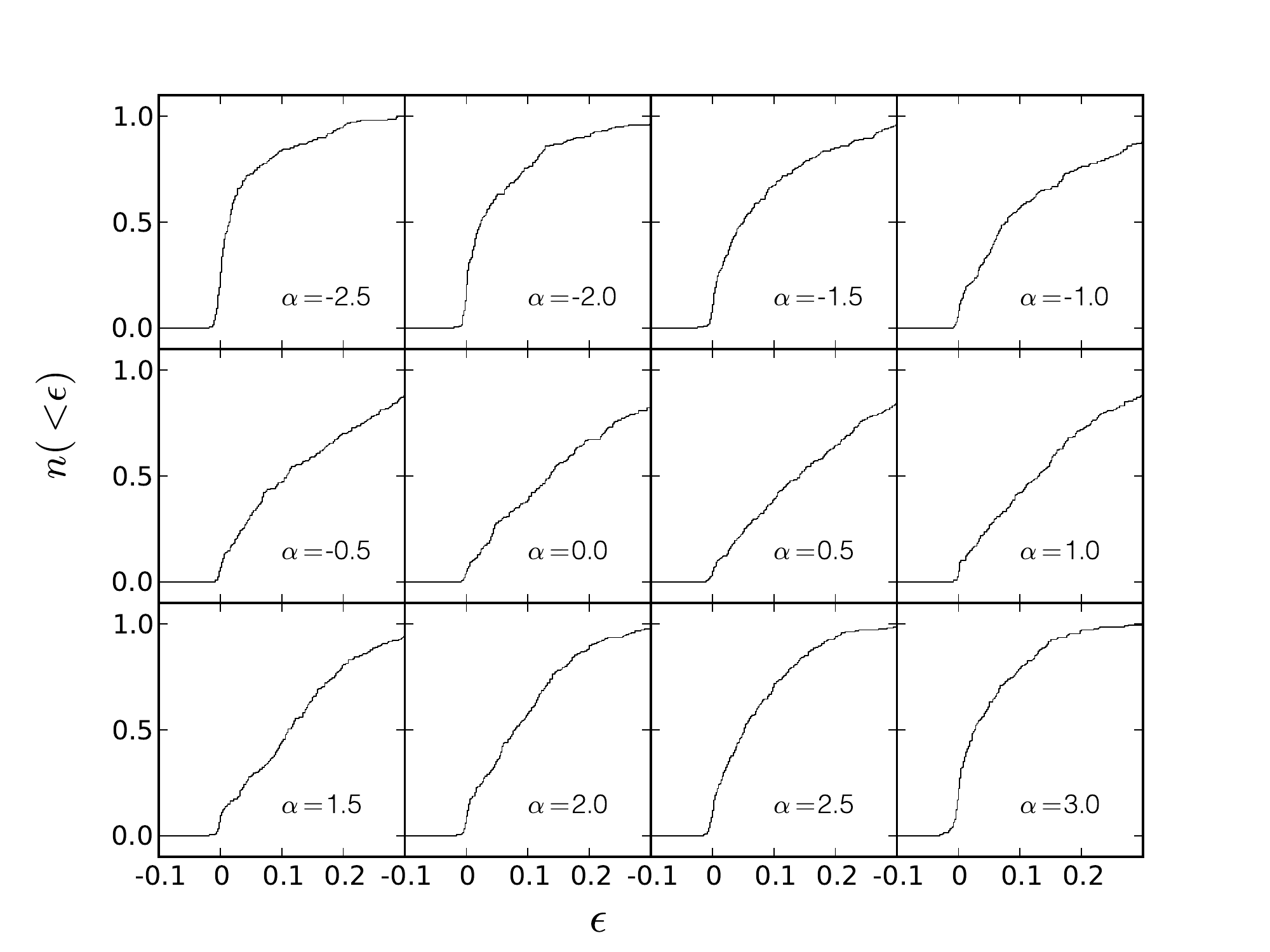}
\caption[$P_2/P_1$ for all $\alpha$]{
Cumulative histogram of the final $\epsilon$ as a result of the interactions 
between a planetesimal disk and resonant planet pairs. Each panel shows results 
from models with all $m_1/m_2$ and $m_d/m_p$ but with a particular $\alpha$, 
denoted in the panels. Planetesimal interactions predominantly create $\epsilon > 0$ 
and naturally explains the asymmetry in the period ratio distribution observed in 
the Kepler data, regardless of the local disk profile.
}
\label{fig:alpha}
\end{center}
\end{figure*} 
%
%
%%%%%%%%%%%%%%%%%%%%%%%%%%%%%%%%%%%%%%%%%%%%%%%%
%FIGURE: MIGRATION DIRECTION
%%%%%%%%%%%%%%%%%%%%%%%%%%%%%%%%%%%%%%%%%%%%%%%%
%
\begin{figure}
\begin{center}
\plotone{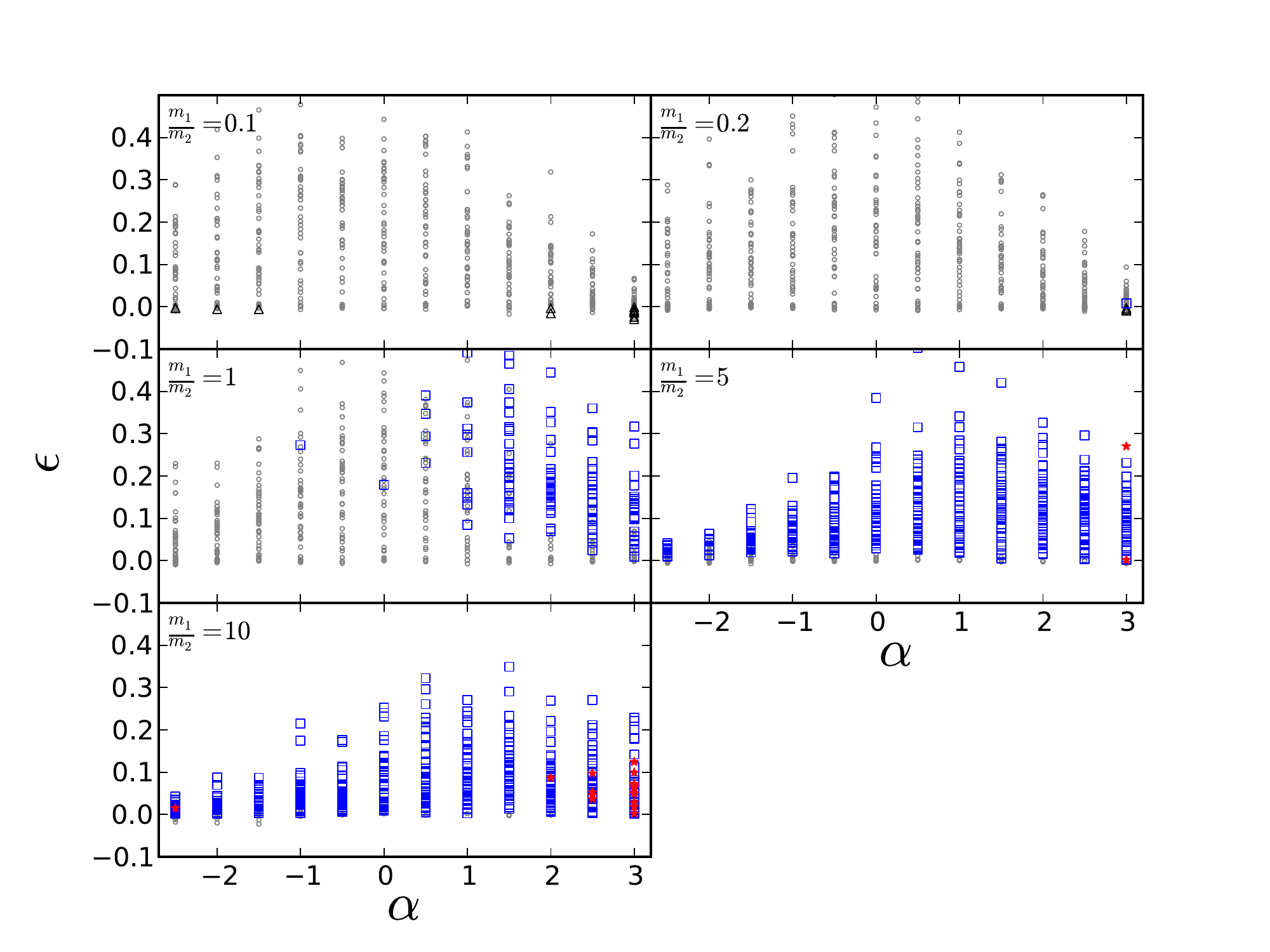}
\caption[Evolution]{
$\epsilon$ vs $\alpha$ for all studied initial $m_1/m_2$, each panel for a given value noted in the 
corresponding panel. Grey dots, black triangles, blue squares, and red stars denote cases where both 
$a_1$ and $a_2$ decreased, $a_1$ increased and $a_2$ decreased, $a_1$ decreased and $a_2$ increased, 
and both $a_1$ and $a_2$ increased after {\tt Stage 3}, respectively. Depending 
on the details of initial properties such as $m_1/m_2$ and $\alpha$ the direction of migration 
varies for the individual planets. However, for sufficiently massive planetesimal 
disks, $\epsilon$ typically grows to significant positive values. 
}
\label{fig:migdir}
\end{center}
\end{figure} 
%

%
%%%%%%%%%%%%%%%%%%%%%%%%%%%%%%%%%%%%%%%%%%%%%%%%
%RESULTS
%%%%%%%%%%%%%%%%%%%%%%%%%%%%%%%%%%%%%%%%%%%%%%%%
\section{Results}
\label{S:results}
Scattering and accretion of the planetesimals perturb the planetary orbits stochastically 
(see Figure\ \ref{fig:evolution}, Right Panels for an example). 
Initially, $\epsilon$ changes chaotically due to individual planet-planetesimal interactions. 
No single interaction is strong enough to break the resonance. These interactions
damp the eccentricities of both planets. After sufficiently large number of interactions 
the overall perturbations disrupt the resonance indicated by recirculation of the resonance 
angles. Once the resonance is disrupted, further interactions steadily increase $\epsilon$ 
until there are insufficient planetesimals left in the disk that can interact with the planets 
(e.g., Figure\ \ref{fig:profile}). 
Then $\epsilon$ reaches a steady value, $\sim 0.1$ in this example (Figure\ \ref{fig:evolution}).  

For a representative set of our models ($m_d/m_p=0.5$, $\alpha=-1.5$, $0$, $1.5$, and $m_1/m_2=0.1$, $1$, $10$) 
we turn on extensive logging to track close encounters, defined as planetesimals coming within 
$1$ Hill radius ($R_H$) of a planet.   
Outcomes of each planet-planetesimal close encounter 
depends on the details of the initial conditions including $m_1/m_2$ and 
$\alpha$ (Figure\ \ref{fig:cewhich}). For our choice of planet properties (mass and density), planetesimals 
from a large range in initial positions ($0.3\lesssim a_{\rm{ini}}/\au\lesssim1$) come within $1R_H$ of both planets. 
Planetesimals that come within $1R_H$ of either the inner or the outer planet had $a_{\rm{ini}}$ predominantly 
inside of the inner planet or outside of the outer planet. The fraction of all planetesimal orbits at the beginning 
of \stagethree\ that come within $1R_H$ of a planet varies widely depending on $m_1/m_2$ and $\alpha$. For 
example, $\approx6\%$ of all planetesimals come within $1R_H$ of a planet during the \stagethree\ evolution 
for one model with $m_1/m_2=1$, $\alpha=1.5$, $m_d/m_p=0.5$, whereas, this fraction is $\approx50\%$ 
for another model with $m_1/m_2=0.1$, $\alpha=-1.5$ and $m_d/m_p=0.5$. The fraction of planetesimals taking part 
in close encounters also vary by large fractions from model to model with the same initial $m_1/m_2$, 
$\alpha$, and $m_d/m_p$ but different simply by random realization of the initial conditions of planetesimals orbits 
(e.g., $5\%$ and $12\%$ between two models with $m_1/m_2=1$, $\alpha=1.5$, and $m_d/m_p=0.5$).   
The ratio of the number of planetesimals that has at least one close encounter with the planets and the 
number of planetesimals that physically collide with one of the planets also can vary widely ($\approx20\%$ 
for one of our models with $m_1/m_2=1$, $\alpha=-1.5$, $m_d/m_p=0.5$, and $40\%$ for $m_1/m_2=10$, 
$\alpha=-1.5$, $m_d/m_p=0.5$). 
The fraction of accreted planetesimals that had close encounters with the planets 
prior to accretion varies stochastically between realizations with the same initial values 
for $m_1/m_2$, $\alpha$, and $m_d/m_p$. 
For example, two different models both with $m_1/m_2=1$, $\alpha=-1.5$ and $m_d/m_p=0.5$ 
show $24\%$ and $33\%$ for this quantity. Planetesimals that finally collide with one of the planets, often 
had multiple close encounters with both planets prior to accretion. Among models with extensive logging of this 
information we find that the ratio of the number of accreted planetesimals that had multiple close encounters with 
both planets before they were finally accreted and the number of all accreted planetesimals is typically large and varies 
between $41\%$ (for a model with $m_1/m_2=0.1$, $\alpha=-1.5$, $m_d/m_p=0.5$) to $65\%$ (for a model with 
$m_1/m_2=1$, $\alpha=-1.5$, $m_d/m_p=0.5$).     

The quantitative results 
including when the resonance is broken and the final value of $\epsilon$ vary from 
model to model. However, the qualitative nature of the evolution for the planetary orbits remains 
the same for all 
$m_1/m_2$, $\alpha$, and $m_d/m_p$ we have considered. For all models with a sufficiently 
massive initial planetesimal disk ($m_d/m_p\geqslant0.2$) the resonance is typically broken 
and $\epsilon$ grows 
to positive values for all $\alpha$ (Figure\ \ref{fig:alpha}). These results indicate 
that the observed asymmetry in the $\epsilon$-distribution for near-resonance planet pairs 
in the \kepler\ data, namely, an overabundance of planet pairs just wide of the 
$2\colon1$ and $3:2$ MMRs and a dearth of systems narrow of them, can be a natural outcome 
of interactions between resonant planet pairs and planetesimals from a planetesimal 
disk. This basic outcome is insensitive to the details of the values for $\alpha$, $m_1/m_2$, and 
$m_d/m_p$ as long as $m_d/m_p$ is sufficiently high.  

For a given $m_d/m_p$, disks with flatter profiles contain a larger number of 
planetesimals close to the planetary orbits compared to steeper profiles 
because of our numerical setup that ensures that 
the disk edges are always sufficiently far from the planetary orbits (\S\ref{S:stage2}). 
As a result, the shape of the final $\epsilon$ distribution depends on $\alpha$ 
(Figure\ \ref{fig:alpha}), primarily because $\alpha$ affects the number of planetesimals 
that can interact with the planets. 
A flatter disk profile ($|\alpha|<1$) tends to create larger $\epsilon$ 
values overall. The extreme positive and negative $\alpha$ models result in very similar 
distributions for the final $\epsilon$ values. 

The migration directions for individual planets 
do depend on the details of these initial properties 
(Figure\ \ref{fig:migdir}). We first focus on systems where the resonance is broken and the final 
$\epsilon$ has grown significantly ($\gtrsim0.01$). Except for a few cases with $m_1/m_2>1$ and 
$\alpha>2$ the inner planet always migrates inwards. For $m_1/m_2<1$, for all $\alpha$, the outer planet 
also moves inwards but at a smaller rate compared to the inner planet such that the orbits diverge and 
$\epsilon$ values grow. In contrast, for $m_1/m_2>1$ the outer planet migrates outwards for all $\alpha$. 
For $m_1/m_2=1$, both outcomes are possible depending on $\alpha$. If $\alpha>1$, then planetary orbits 
diverge via inner planet moving in and outer planet moving out. Otherwise both planets move in. 
For the cases where $\epsilon$ stays small (e.g., when the resonance is not broken as a result of the 
planet-planetesimal interactions), $\epsilon$ can have both positive and negative values and both planets 
may move slightly in or out. We also find that for any given $\alpha$ the dispersion in the growth of $\epsilon$ 
is typically higher if $m_1/m_2<1$. 
  
%
%%%%%%%%%%%%%%%%%%%%%%%%%%%%%%%%%%%%%%%%%%%%%%%%
%FIGURE: MDOVERMP AND PRATIO PERCENTILES
%%%%%%%%%%%%%%%%%%%%%%%%%%%%%%%%%%%%%%%%%%%%%%%%
%
\begin{figure}
\begin{center}
\plotone{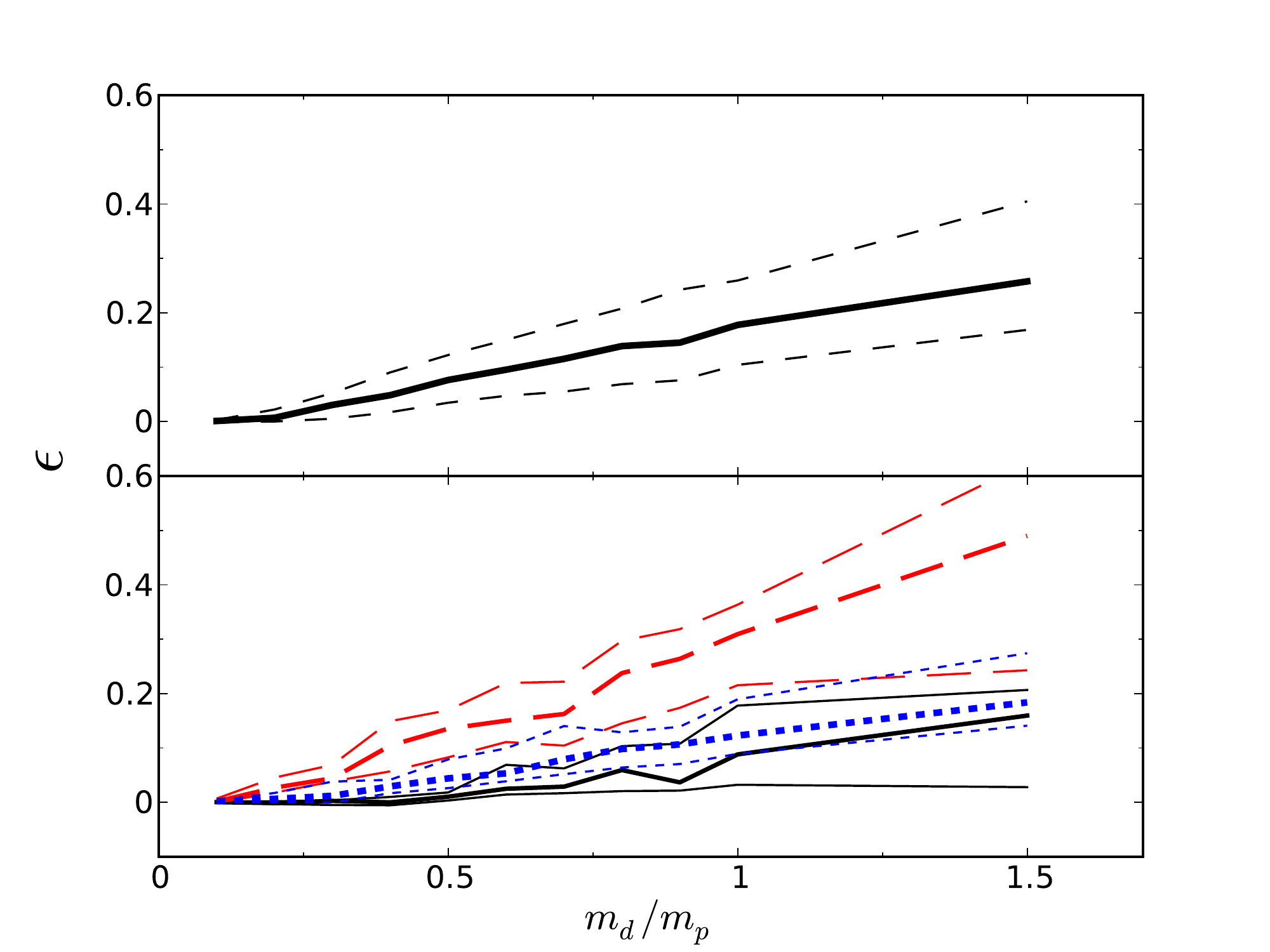}
\caption[$\epsilon$ percentiles vs $m_d/m_p$]{
Quartile values for the $\epsilon$-distributions as a function of $m_d/m_p$. 
The top panel shows results from all our models. The bottom panel shows results 
from models with three specific planetesimals disk profiles $\alpha = -2.5$ (black solid), 
$0$ (red long-dashed), and $2.5$ (blue short-dashed). In each panel, the thick 
and thin lines show the median, and the 25\%(below) and 75\% (above) percentiles. 
The quartiles of the $\epsilon$ distributions increase almost linearly with $m_d/m_p$. 
The slope and the minimum value for disruption of resonance and growth of $\epsilon$ 
depends on $\alpha$. 
}
\label{fig:mdovermp}
\end{center}
\end{figure} 
\begin{figure}
\begin{center}
\plotone{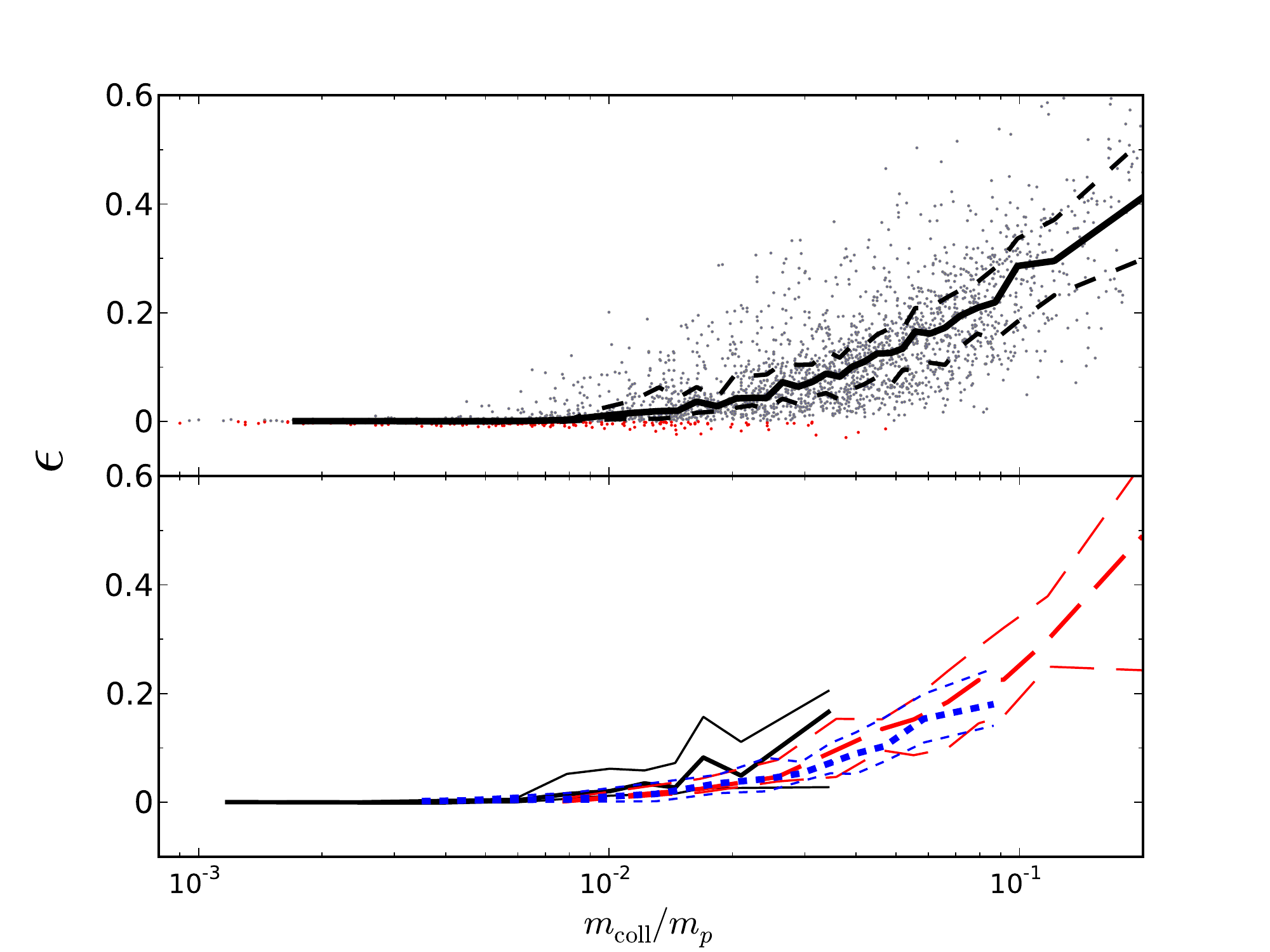}
\caption[$\epsilon$ percentiles vs $\mcoll/m_p$]{
Same as Figure\ \ref{fig:mdovermp} but here we show quartile values for the $\epsilon$ 
distributions as a function of $\mcoll/m_p$. 
The grey and red dots in the top panel show the final $\epsilon > 0$ and $\epsilon \geq 0$ 
values, respectively. The quartiles of the $\epsilon$-distributions increase almost 
linearly both with $\mcoll/m_p$, similar to $m_d/m_p$. Clear threshold value 
for $\mcoll/m_p\sim0.008$ can be seen for disruption of resonance and growth of 
$\epsilon$. The rate of increase of the quartiles as a function of $\mcoll/m_p$ does not 
show a strong dependence on $\alpha$. 
}
\label{fig:mcollovermp}
\end{center}
\end{figure} 

Figure\ \ref{fig:mdovermp} shows the dependence of the final $\epsilon$ on $m_d/m_p$. 
The quartiles (25\%, 50\%, and 75\%) for the final $\epsilon$ 
distributions increase almost linearly with increasing $m_d/m_p$ (Top-Left). 
For $m_d/m_p\leqslant0.2$, the resonance is typically not disrupted and $\epsilon$ remains small 
($\sim 10^{-3}$). The critical value for $m_d/m_p$ to break the $2\colon1$ resonance and 
drive up $\epsilon$ depends on $\alpha$ because the number of planetesimals close 
enough to interact with the planets depends on $\alpha$. 
For example, for models with $\alpha=-2.5$ all quartiles for the $\epsilon$-distribution 
remains small for $m_d/m_p\lesssim0.5$. 
The slope of the increase of the quartiles 
with respect to $m_d/m_p$ also stays relatively low. For another steep disk profile $\alpha=2.5$ 
the results are quite similar, and show only a small difference in the exact value of the 
critical $m_d/m_p$ for resonance disruption and a modest increase of the slope. 
In contrast, for a flat disk profile $\alpha=0$ we find that the critical $m_d/m_p$ value 
is reduced significantly and the quartiles of the $\epsilon$ distribution show a much 
steeper slope. These differences largely originate from the difference in the number of 
planetesimals close enough to interact with the planet pairs depending on $\alpha$. 
Indeed Figure\ \ref{fig:cewhich} shows that flat ($\alpha=0$) disk profiles result in many more 
planet-planetesimal close encounters compared to steeper profiles (i.e., $\alpha=1.5, -1.5$). 
For the steeper disk profiles a large number of planetesimals do not suffer any strong interactions 
with the planets simply because they are 
too far from the planets. Thus, the critical 
value for $m_d/m_p$ remains dependent, on the initial power law profile and the 
locations of the disk edges relative to the planetary orbits.  

While $m_d/m_p$ provides important information for this process such as how much total 
angular momentum or energy must be stored in the planetesimal orbits (for a given choice of planetesimal 
boundaries and $\alpha$), the quantitative results 
such as how many planetesimals take part in strong interaction with the planets depend on the 
choices of the disk profile and the disk edges. For example, if the planetesimal disk spans a smaller 
region, same degree of planet-planetesimal interactions may be achieved from a lower $m_d/m_p$.  
A less ambiguous and more dynamically informative quantity is 
the total mass of planetesimals that had strong interactions with the planets ($\mint$). 

As a proxy to strong interactions we calculate the total mass of planetesimals 
accreted by the planets ($\mcoll$). 
Using the representative models with extensive logging of all close encounters 
(Figure\ \ref{fig:cewhich}) we find that $\mcoll$ (equivalently number of collisions) 
is highly correlated with $\mint$. The 
Pearson's correlation coefficient between $\mcoll$ and $\mint$ is $0.95$. 
This confirms that $\mcoll$ should serve as a 
good tracer of the total mass of 
dynamically important planetesimals. 
Alternatively, several other quantities 
including the total number of planet-planetesimal strong encounters, 
or total number of planetesimals within regions close to either planets could have been 
used. However, using $\mcoll$ as a tracer of the mass of dynamically 
important planetesimals gives us two advantages. Tracking $\mcoll$ does not require 
extensive logging of close encounters which slows these simulations significantly. 
In addition, physical collisions may lead to observable 
differences in atmospheric properties and may create strange mass-radius relations 
for these planets, which make $\mcoll$ a very interesting quantity.

The quartiles for the final $\epsilon$-distributions increase almost linearly with increasing 
$\mcoll/m_p$, similar to what we see in the case of $m_d/m_p$ 
(Figure\ \ref{fig:mcollovermp}). 
However, we find that the slopes for the quartiles of $\epsilon$ distributions with respect 
to $\mcoll/m_p$ do not vary as much with changing $\alpha$ as was seen for $m_d/m_p$. 
Several key factors related to the inherent nature of this process are apparent 
from the dependence of $\epsilon$ on $\mcoll/m_p$.  
Each planet-planetesimal interaction results in a small change in $a$, $\delta a<\Delta a$, 
the resonance width. The qualitative behavior of each random interaction 
between a planetesimal and one of the planets is similar to a random 
walk (bottom right panel, Figure 1) with step size $\delta a$. In this picture, higher 
$\mcoll/m_p$ means larger number of steps. If $\mcoll/m_p$ is not sufficiently high, 
these random steps are unlikely to drive the planets out of resonance. For a sufficiently 
large $\mcoll/m_p$, the cumulative effect of several interactions may result in a 
overall change $>\Delta a$ disrupting the resonance. Afterwards, planets migrate 
further apart since they are able to interact with more planetesimals near their new 
positions. The inherent stochastic nature of this process is manifested by the fact that 
for any given $\mcoll/m_p$, $\epsilon$ can attain a range of final values. Even when 
$\mcoll/m_p$ is sufficiently high, the resonance may not be disrupted 
in a small number of systems. Nevertheless, increase in $\mcoll/m_p$ 
increases the probability of disruption of resonance and growth of $\epsilon$.   

We find that the threshold for resonance disruption and growth of $\epsilon$ 
is $\mcoll/m_p\approx0.008$, for our choice of planet masses and densities which are guided 
by typical values observed of the \kepler\ systems (\S\ref{S:modeling}). 
If the total mass of 
accreted planetesimals is lower than this critical value, then there is little 
chance for growth of $\epsilon$. 
Such threshold-like behavior naturally explains 
why the positive $\epsilon\sim0.01$--$0.2$ are seen only among much smaller, hence 
less massive \kepler\ systems. The total mass in planetesimals after gas dispersal 
is seldom sufficient to disrupt resonances between the more massive RV-discovered 
planets. Even if giant and small planets, both get trapped in the same MMRs during 
the gas-disk dominated evolution, planetesimal interactions produce significant 
$\epsilon>0$ only for the low-mass planets. 

Due to the stochastic nature of this process, sometimes the resonance is not 
disrupted even if $m_d/m_p$ or $\mcoll/m_p$ are sufficiently high. 
In these cases $\epsilon$ values remain small $\sim 10^{-3}$. 
Moreover, in such cases, planet pairs can attain both positive and negative $\epsilon$ values. 
Negative $\epsilon$ values are always small $|\epsilon|\sim10^{-3}$ 
(Figure\ \ref{fig:mcollovermp}) and are only seen in systems 
where the initial resonance is not disrupted. 

In each model we choose a fixed density for the planets. \kepler\ data, however, show that 
planets of very similar sizes can have widely varying densities 
\citep{2014ApJS..210...20M,2014ApJ...783L...6W,2015ApJ...798L..32C}. Using a few representative models 
we test whether the extent of divergence of planetary orbits via planet-planetesimal interactions depend on 
the planetary densities. 
We find that for the full range in densities 
(point mass to $0.5\,\gcc$) 
planetary orbits diverge from the initial resonant configuration and $\epsilon$ can attain 
large ($\geq0.01$) positive values. For example, multiple realizations of 
models with $m_1/m_2=1$, $\alpha=-1.5$, and $m_d/m_p=0.5$ show final $\epsilon$ values ranging between 
$0.06$ to $0.14$ for the point-mass planets, $0.07$ to $0.11$ for $\rho_p=1.68\,\gcc$, and $0.05$ to $0.09$ 
for $\rho_p=0.5\,\gcc$. Hence, we conclude that growth of $\epsilon$ via planetesimal scattering 
should not critically depend on the choice of 
planetary densities. 

The extent of planetesimal driven migration may depend on the initial $e$ of the 
planet orbits and the RMS $e$ of planetesimals in \stagethree\ 
\citep[\S\ref{S:stage2},\ref{S:stage3};][]{2009Icar..199..197K,2014Icar..232..118M}. 
Effects of the $e$-distribution in planetesimal orbits on the resonant planets' migration 
may be interesting to explore in a future study.  

%
%%%%%%%%%%%%%%%%%%%%%%%%%%%%%%%%%%%%%%%%%%%%%%%%
%FIGURE: 2D HISTOGRAMS MCOLL AND ALPHA
%%%%%%%%%%%%%%%%%%%%%%%%%%%%%%%%%%%%%%%%%%%%%%%%
%
%
\begin{figure}
\begin{center}
\plotone{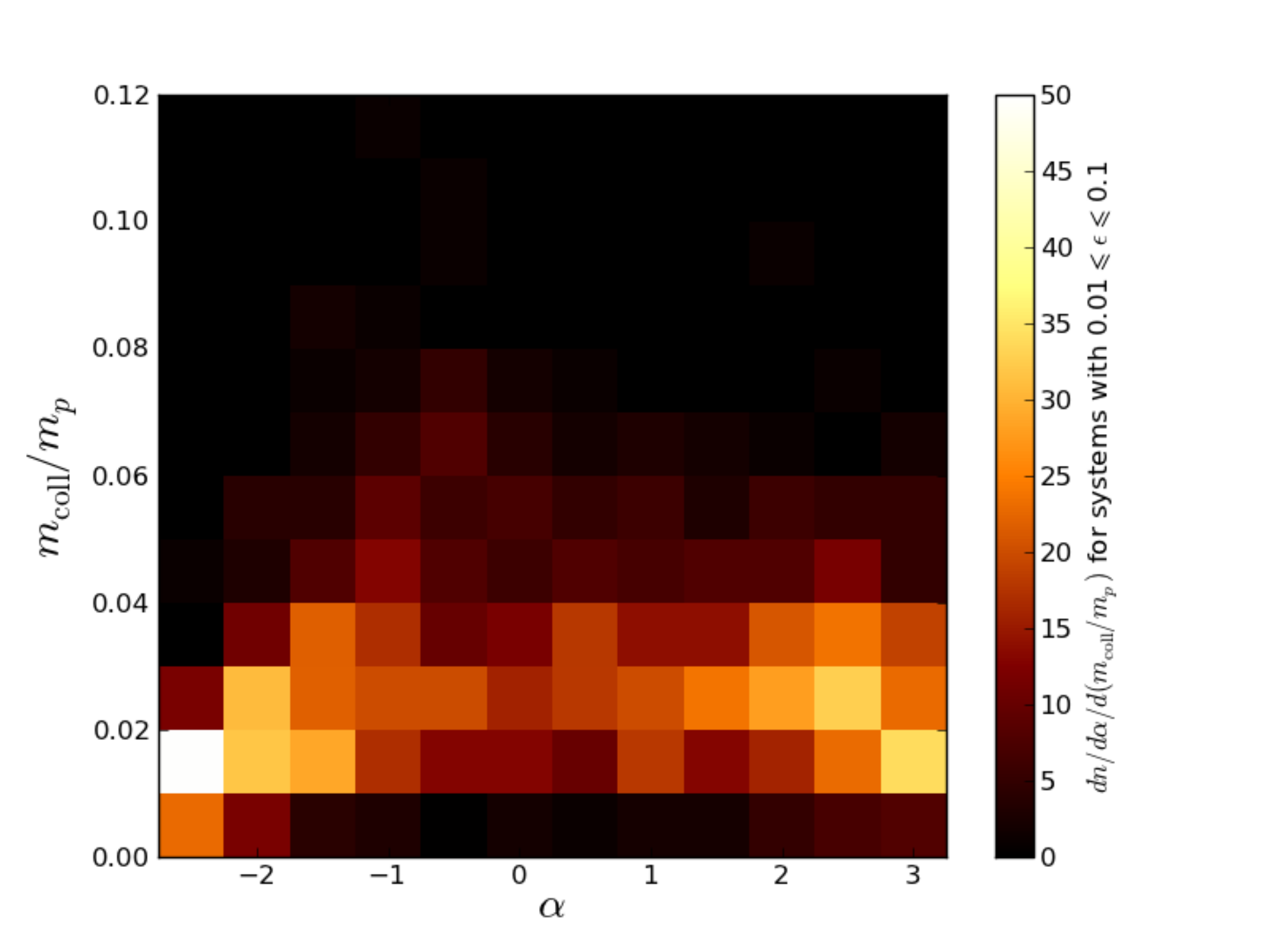}
\caption[2D histogram]{
Two dimensional histogram of the number of systems with final 
$0.01\leq\epsilon\leq0.1$ as a function of $\alpha$ and $\mcoll/m_p$. The colors 
denote the number of systems in each two dimensional bin $dn/d\alpha/d(\mcoll/m_p)$. 
Typically, flatter disks show a larger $\mcoll/m_p$ range compared to steeper disks 
for creation of $\epsilon$ values in the same range.
}
\label{fig:2dfig}
\end{center}
\end{figure} 

Our results indicate that the near-resonant planet pairs in the Kepler data can 
provide hitherto unprecedented constraints on the structure and mass of the 
exo-planetesimal disks as the system emerges from the gas disk. For many systems 
with a pair of near-resonant planets, planet masses can be measured using analysis 
of transit time variations 
\citep[e.g.,][]{2012ApJ...756..186S,2012ApJ...756..185F,2013MNRAS.428.1077S,2013ApJS..208...16M,2014ApJ...787...80H,2014ApJS..210...25X}. 
Observational 
constraints on the planet masses and $\epsilon$ for a particular system can lead 
to constraints on the total mass and the surface density profile of the planetesimal disk 
when the gas dissipates, assuming that the $\epsilon$ is generated via 
planet-planetesimal-disk interactions using, e.g., Figure\ \ref{fig:2dfig}. 
For example, to create the observed $\epsilon=0.18$ for the 
\kepler-18bc system, $0.02\leq\mcoll/m_p\leq0.14$ is required for a 
reasonable range $-2.5\leq\alpha\leq1$. Disks with steeper density profiles show a 
smaller range in $\mcoll/m_p$ to produce the same observed $\epsilon$ 
(Figure\ \ref{fig:2dfig}) compared to flatter disk profiles. 

%
%%%%%%%%%%%%%%%%%%%%%%%%%%%%%%%%%%%%%%%%%%%%%
%DISCUSSION
%%%%%%%%%%%%%%%%%%%%%%%%%%%%%%%%%%%%%%%%%%%%%
%
\section{Discussion}
\label{S:discussion}
In this study we propose a mechanism that can naturally change orbits of planet pairs 
initially trapped in a MMR and create $\epsilon\sim0.01$--$0.2$ (Figure\ \ref{fig:evolution}), 
typical of the observed KPC pairs 
near major MMRs \citep{2011ApJS..197....8L,2014ApJ...790..146F,2014arXiv1409.3320S}. 
Planetary systems with several planets accompanied by a large number of 
planetesimals are thought to form naturally via the core accretion paradigm of planet formation 
\citep[e.g.,][]{2004ARA&A..42..549G,2013apf..book.....A}. 
Planetesimal disks are also seen in the Solar system. 
Planet formation theories also predict trapping of planet pairs via smooth gas-disk driven 
migration. Hence, our adopted initial setup of resonant planet pairs close to planetesimal 
disks after gas dispersal is expected to be common based on the predictions of standard 
planet formation theories. 
We find that interactions between resonant planet pairs and planetesimals in a planetesimal 
disk naturally creates large positive $\epsilon$ if the total mass of accreted planetesimals 
is sufficiently high (Figure\ \ref{fig:mcollovermp}). These interactions also naturally produce 
the observed asymmetry in the $\epsilon$ distribution in the \kepler\ systems for large 
ranges in $\alpha$ (Figure\ \ref{fig:alpha}). 
The existence of a critical value of $\mcoll/m_p$ (Figure\ \ref{fig:mcollovermp})
for resonance disruption and growth of $\epsilon$ also explains why this asymmetry 
is only observed in \kepler\ systems that are typically less massive compared to the 
RV-discovered systems. 

We conducted pure $N$-body integrations using initial conditions generated 
to mimic systems that emerged from a dissipative gas disk. To more rigorously ascertain the initial conditions 
for resonant planet pairs and residual planetesimal disks as they emerge 
out of a gas disk together, one would need to include gas, planetesimals and planets in a self consistent model. 
This is extremely challenging and beyond the scope of our study. However, since the basic result of 
$\epsilon$-growth seems to be common for such a wide range of $\alpha$ (even extreme 
positive values), these conclusions may 
not be sensitive to the details of the properties of the planetesimals disk. The only way 
to avoid our proposed mechanism would be if resonant planet pairs never are near significantly 
massive planetesimals disks when disk disperses.    

If indeed planet-planetesimals interactions are responsible for the observed asymmetric 
distribution of $\epsilon$ for near resonant planet pairs in the \kepler\ data, then 
the high frequency of large-$\epsilon$, near-resonant 
\kepler\ systems indicates that planet-planetesimal interactions after gas-disk 
dissipation is an important evolutionary stage for all small planets independent 
of their proximity to a resonance. Collisions of planetesimals of total mass $\sim$ 
few percent of the planet mass is often needed for the observed high values of $\epsilon$ 
for \kepler-discovered near-resonance planet pairs. Many small \kepler\ planets 
are measured to have surprisingly low $\rho_p$, which makes collisions even more efficient for these 
planets. 
These collisions 
may affect the 
planets' atmospheric properties and compositions especially for the low-mass atmospheres 
believed to be typical for small planets \citep{2014arXiv1407.4457R}. 
Such accretions could also potentially create an anomalous 
mass-radius relationship for near-resonant low-mass planets. Therefore, we 
encourage detailed modeling of the effects of planetesimal collisions on the 
atmospheric properties of these planets, especially for pairs with large observed $\epsilon$ 
values.

%
%%%%%%%%%%%%%%%%%%%%%%%%%%%%%%%%%%%%%%%%%%%%%%%%
%ACKNOWLEDGMENT
%%%%%%%%%%%%%%%%%%%%%%%%%%%%%%%%%%%%%%%%%%%%%%%%
%
\acknowledgments We thank the referee for his detailed report and constructive 
comments. This research was supported by NASA Origins of Solar 
Systems awards NNX09AB35G and NNX13AF61G, NASA Applied Information 
Systems Research Program award NXX09AM41G, Kepler Participating Scientist 
Program award NNX12AF73Gm, the University of Florida and the Pennsylvania 
State University Center for Exoplanets and Habitable Worlds. S.C. also acknowledges 
support from University of Florida theory postdoctoral fellowship and CIERA fellowship 
at Northwestern University. The authors 
acknowledge the University of Florida High-Performance Computing Center for 
providing computational resources and support that have contributed to the research 
results reported within this paper.

%\clearpage
%\nocite{*}
%\bibliography{biblio_planetesimal}

\begin{thebibliography}{51}
\expandafter\ifx\csname natexlab\endcsname\relax\def\natexlab#1{#1}\fi

\bibitem[{{Armitage}(2013)}]{2013apf..book.....A}
{Armitage}, P.~J. 2013, {Astrophysics of Planet Formation}

\bibitem[{{Batalha} {et~al.}(2013){Batalha}, {Rowe}, {Bryson}, {Barclay},
  {Burke}, {Caldwell}, {Christiansen}, {Mullally}, {Thompson}, {Brown},
  {Dupree}, {Fabrycky}, {Ford}, {Fortney}, {Gilliland}, {Isaacson}, {Latham},
  {Marcy}, {Quinn}, {Ragozzine}, {Shporer}, {Borucki}, {Ciardi}, {Gautier},
  {Haas}, {Jenkins}, {Koch}, {Lissauer}, {Rapin}, {Basri}, {Boss}, {Buchhave},
  {Carter}, {Charbonneau}, {Christensen-Dalsgaard}, {Clarke}, {Cochran},
  {Demory}, {Desert}, {Devore}, {Doyle}, {Esquerdo}, {Everett}, {Fressin},
  {Geary}, {Girouard}, {Gould}, {Hall}, {Holman}, {Howard}, {Howell},
  {Ibrahim}, {Kinemuchi}, {Kjeldsen}, {Klaus}, {Li}, {Lucas}, {Meibom},
  {Morris}, {Pr{\v s}a}, {Quintana}, {Sanderfer}, {Sasselov}, {Seader},
  {Smith}, {Steffen}, {Still}, {Stumpe}, {Tarter}, {Tenenbaum}, {Torres},
  {Twicken}, {Uddin}, {Van Cleve}, {Walkowicz}, \&
  {Welsh}}]{2013ApJS..204...24B}
{Batalha}, N.~M., {Rowe}, J.~F., {Bryson}, et~al. 2013, \apjs, 204, 24

\bibitem[{{Batygin} \& {Morbidelli}(2013)}]{2013AJ....145....1B}
{Batygin}, K. \& {Morbidelli}, A. 2013, \aj, 145, 1

\bibitem[{{Beaug{\'e}} \& {Nesvorn{\'y}}(2012)}]{2012ApJ...751..119B}
{Beaug{\'e}}, C. \& {Nesvorn{\'y}}, D. 2012, \apj, 751, 119

\bibitem[{{Borucki} {et~al.}(2010){Borucki}, {Koch}, {Basri}, {Batalha},
  {Brown}, {Caldwell}, {Caldwell}, {Christensen-Dalsgaard}, {Cochran},
  {DeVore}, {Dunham}, {Dupree}, {Gautier}, {Geary}, {Gilliland}, {Gould},
  {Howell}, {Jenkins}, {Kondo}, {Latham}, {Marcy}, {Meibom}, {Kjeldsen},
  {Lissauer}, {Monet}, {Morrison}, {Sasselov}, {Tarter}, {Boss}, {Brownlee},
  {Owen}, {Buzasi}, {Charbonneau}, {Doyle}, {Fortney}, {Ford}, {Holman},
  {Seager}, {Steffen}, {Welsh}, {Rowe}, {Anderson}, {Buchhave}, {Ciardi},
  {Walkowicz}, {Sherry}, {Horch}, {Isaacson}, {Everett}, {Fischer}, {Torres},
  {Johnson}, {Endl}, {MacQueen}, {Bryson}, {Dotson}, {Haas}, {Kolodziejczak},
  {Van Cleve}, {Chandrasekaran}, {Twicken}, {Quintana}, {Clarke}, {Allen},
  {Li}, {Wu}, {Tenenbaum}, {Verner}, {Bruhweiler}, {Barnes}, \&
  {Prsa}}]{2010Sci...327..977B}
{Borucki}, W.~J., {Koch}, D., {Basri}, et~al. 2010, Science, 327, 977

\bibitem[{{Borucki} {et~al.}(2011){Borucki}, {Koch}, {Basri}, {Batalha},
  {Brown}, {Bryson}, {Caldwell}, {Christensen-Dalsgaard}, {Cochran}, {DeVore},
  {Dunham}, {Gautier}, {Geary}, {Gilliland}, {Gould}, {Howell}, {Jenkins},
  {Latham}, {Lissauer}, {Marcy}, {Rowe}, {Sasselov}, {Boss}, {Charbonneau},
  {Ciardi}, {Doyle}, {Dupree}, {Ford}, {Fortney}, {Holman}, {Seager},
  {Steffen}, {Tarter}, {Welsh}, {Allen}, {Buchhave}, {Christiansen}, {Clarke},
  {Das}, {D{\'e}sert}, {Endl}, {Fabrycky}, {Fressin}, {Haas}, {Horch},
  {Howard}, {Isaacson}, {Kjeldsen}, {Kolodziejczak}, {Kulesa}, {Li}, {Lucas},
  {Machalek}, {McCarthy}, {MacQueen}, {Meibom}, {Miquel}, {Prsa}, {Quinn},
  {Quintana}, {Ragozzine}, {Sherry}, {Shporer}, {Tenenbaum}, {Torres},
  {Twicken}, {Van Cleve}, {Walkowicz}, {Witteborn}, \&
  {Still}}]{2011ApJ...736...19B}
{Borucki}, W.~J., {Koch}, D.~G., {Basri}, G., et~al. 2011,
  \apj, 736, 19

\bibitem[{{Bromley} \& {Kenyon}(2011)}]{2011ApJ...735...29B}
{Bromley}, B.~C. \& {Kenyon}, S.~J. 2011, \apj, 735, 29

\bibitem[{{Bryden} {et~al.}(2000){Bryden}, {R{\'o}{\.z}yczka}, {Lin}, \&
  {Bodenheimer}}]{2000ApJ...540.1091B}
{Bryden}, G., {R{\'o}{\.z}yczka}, M., {Lin}, D.~N.~C., \& {Bodenheimer}, P.
  2000, \apj, 540, 1091

\bibitem[{{Burke} {et~al.}(2014){Burke}, {Bryson}, {Mullally}, {Rowe},
  {Christiansen}, {Thompson}, {Coughlin}, {Haas}, {Batalha}, {Caldwell},
  {Jenkins}, {Still}, {Barclay}, {Borucki}, {Chaplin}, {Ciardi}, {Clarke},
  {Cochran}, {Demory}, {Esquerdo}, {Gautier}, {Gilliland}, {Girouard}, {Havel},
  {Henze}, {Howell}, {Huber}, {Latham}, {Li}, {Morehead}, {Morton}, {Pepper},
  {Quintana}, {Ragozzine}, {Seader}, {Shah}, {Shporer}, {Tenenbaum}, {Twicken},
  \& {Wolfgang}}]{2014ApJS..210...19B}
{Burke}, C.~J., {Bryson}, S.~T., {Mullally}, F., et~al. 2014,
  \apjs, 210, 19

\bibitem[{{Butler} {et~al.}(2006){Butler}, {Wright}, {Marcy}, {Fischer},
  {Vogt}, {Tinney}, {Jones}, {Carter}, {Johnson}, {McCarthy}, \&
  {Penny}}]{2006ApJ...646..505B}
{Butler}, R.~P., {Wright}, J.~T., {Marcy}, G.~W., et~al. 2006, \apj, 646, 505

\bibitem[{{Chambers}(1999)}]{1999MNRAS.304..793C}
{Chambers}, J.~E. 1999, \mnras, 304, 793

\bibitem[{{Chatterjee} {et~al.}(2008){Chatterjee}, {Ford}, {Matsumura}, \&
  {Rasio}}]{2008ApJ...686..580C}
{Chatterjee}, S., {Ford}, E.~B., {Matsumura}, S., \& {Rasio}, F.~A. 2008, \apj,
  686, 580

\bibitem[{{Chatterjee} \& {Tan}(2014)}]{2014ApJ...780...53C}
{Chatterjee}, S. \& {Tan}, J.~C. 2014, \apj, 780, 53

%\bibitem[{{Chatterjee} \& {Tan}(2014{\natexlab{b}})}]{2014arXiv1411.2629C}
%---. 2014{\natexlab{b}}, arXiv:1411.2629

\bibitem[{{Chatterjee} \& {Tan}(2015)}]{2015ApJ...798L..32C}
{Chatterjee}, S. \& {Tan}, J.~C. 2015, \apjl, 798, 32

\bibitem[{{Chiang} \& {Laughlin}(2013)}]{2013MNRAS.431.3444C}
{Chiang}, E. \& {Laughlin}, G. 2013, \mnras, 431, 3444

\bibitem[{{Delisle} \& {Laskar}(2014)}]{2014A&A...570L...7D}
{Delisle}, J.~B. \& {Laskar}, J. 2014, \aap, 570, L7

\bibitem[{{Fabrycky} {et~al.}(2014){Fabrycky}, {Lissauer}, {Ragozzine}, {Rowe},
  {Steffen}, {Agol}, {Barclay}, {Batalha}, {Borucki}, {Ciardi}, {Ford},
  {Gautier}, {Geary}, {Holman}, {Jenkins}, {Li}, {Morehead}, {Morris},
  {Shporer}, {Smith}, {Still}, \& {Van Cleve}}]{2014ApJ...790..146F}
{Fabrycky}, D.~C., {Lissauer}, J.~J., {Ragozzine}, D., et~al. 2014, \apj,
  790, 146

\bibitem[{{Fang} \& {Margot}(2012)}]{2012ApJ...761...92F}
{Fang}, J. \& {Margot}, J.-L. 2012, \apj, 761, 92

\bibitem[{{Fernandez} \& {Ip}(1984)}]{1984Icar...58..109F}
{Fernandez}, J.~A. \& {Ip}, W.-H. 1984, \icarus, 58, 109

\bibitem[{{Ford} {et~al.}(2012){Ford}, {Ragozzine}, {Rowe}, {Steffen},
  {Barclay}, {Batalha}, {Borucki}, {Bryson}, {Caldwell}, {Fabrycky}, {Gautier},
  {Holman}, {Ibrahim}, {Kjeldsen}, {Kinemuchi}, {Koch}, {Lissauer}, {Still},
  {Tenenbaum}, {Uddin}, \& {Welsh}}]{2012ApJ...756..185F}
{Ford}, E.~B., {Ragozzine}, D., {Rowe}, J.~F., et~al. 2012, \apj, 756, 185

\bibitem[{{Goldreich} {et~al.}(2004){Goldreich}, {Lithwick}, \&
  {Sari}}]{2004ARA&A..42..549G}
{Goldreich}, P., {Lithwick}, Y., \& {Sari}, R. 2004, \araa, 42, 549

\bibitem[{{Goldreich} \& {Tremaine}(1980)}]{1980ApJ...241..425G}
{Goldreich}, P. \& {Tremaine}, S. 1980, \apj, 241, 425

\bibitem[{{Hadden} \& {Lithwick}(2014)}]{2014ApJ...787...80H}
{Hadden}, S. \& {Lithwick}, Y. 2014, \apj, 787, 80

\bibitem[{{Hahn} \& {Malhotra}(1999)}]{1999AJ....117.3041H}
{Hahn}, J.~M. \& {Malhotra}, R. 1999, \aj, 117, 3041

\bibitem[{{Hansen} \& {Murray}(2012)}]{2012ApJ...751..158H}
{Hansen}, B.~M.~S. \& {Murray}, N. 2012, \apj, 751, 158

\bibitem[{{Hansen} \& {Murray}(2013)}]{2013ApJ...775...53H}
---. 2013, \apj, 775, 53

\bibitem[{{Howe} {et~al.}(2014){Howe}, {Burrows}, \&
  {Verne}}]{2014ApJ...787..173H}
{Howe}, A.~R., {Burrows}, A., \& {Verne}, W. 2014, \apj, 787, 173

\bibitem[{{Juri{\'c}} \& {Tremaine}(2008)}]{2008ApJ...686..603J}
{Juri{\'c}}, M. \& {Tremaine}, S. 2008, \apj, 686, 603

\bibitem[{{Kirsh} {et~al.}(2009){Kirsh}, {Duncan}, {Brasser}, \&
  {Levison}}]{2009Icar..199..197K}
{Kirsh}, D.~R., {Duncan}, M., {Brasser}, R., \& {Levison}, H.~F. 2009, \icarus,
  199, 197

\bibitem[{{Lee} {et~al.}(2013){Lee}, {Fabrycky}, \&
  {Lin}}]{2013ApJ...774...52L}
{Lee}, M.~H., {Fabrycky}, D., \& {Lin}, D.~N.~C. 2013, \apj, 774, 52

\bibitem[{{Lee} \& {Peale}(2002)}]{2002ApJ...567..596L}
{Lee}, M.~H. \& {Peale}, S.~J. 2002, \apj, 567, 596

\bibitem[{{Levison} {et~al.}(2007){Levison}, {Morbidelli}, {Gomes}, \&
  {Backman}}]{2007prpl.conf..669L}
{Levison}, H.~F., {Morbidelli}, A., {Gomes}, R., \& {Backman}, D. 2007,
  Protostars and Planets V, 669

\bibitem[{{Lin} {et~al.}(1996){Lin}, {Bodenheimer}, \&
  {Richardson}}]{1996Natur.380..606L}
{Lin}, D.~N.~C., {Bodenheimer}, P., \& {Richardson}, D.~C. 1996, \nat, 380, 606

\bibitem[{{Lissauer} {et~al.}(2011){Lissauer}, {Ragozzine}, {Fabrycky},
  {Steffen}, {Ford}, {Jenkins}, {Shporer}, {Holman}, {Rowe}, {Quintana},
  {Batalha}, {Borucki}, {Bryson}, {Caldwell}, {Carter}, {Ciardi}, {Dunham},
  {Fortney}, {Gautier}, {Howell}, {Koch}, {Latham}, {Marcy}, {Morehead}, \&
  {Sasselov}}]{2011ApJS..197....8L}
{Lissauer}, J.~J., {Ragozzine}, D., {Fabrycky}, D.~C., et~al. 2011,
  \apjs, 197, 8

\bibitem[{{Lithwick} \& {Wu}(2012)}]{2012ApJ...756L..11L}
{Lithwick}, Y. \& {Wu}, Y. 2012, \apjl, 756, L11

\bibitem[{{Marcy} {et~al.}(2014){Marcy}, {Isaacson}, {Howard}, {Rowe},
  {Jenkins}, {Bryson}, {Latham}, {Howell}, {Gautier}, {Batalha}, {Rogers},
  {Ciardi}, {Fischer}, {Gilliland}, {Kjeldsen}, {Christensen-Dalsgaard},
  {Huber}, {Chaplin}, {Basu}, {Buchhave}, {Quinn}, {Borucki}, {Koch}, {Hunter},
  {Caldwell}, {Van Cleve}, {Kolbl}, {Weiss}, {Petigura}, {Seager}, {Morton},
  {Johnson}, {Ballard}, {Burke}, {Cochran}, {Endl}, {MacQueen}, {Everett},
  {Lissauer}, {Ford}, {Torres}, {Fressin}, {Brown}, {Steffen}, {Charbonneau},
  {Basri}, {Sasselov}, {Winn}, {Sanchis-Ojeda}, {Christiansen}, {Adams},
  {Henze}, {Dupree}, {Fabrycky}, {Fortney}, {Tarter}, {Holman}, {Tenenbaum},
  {Shporer}, {Lucas}, {Welsh}, {Orosz}, {Bedding}, {Campante}, {Davies},
  {Elsworth}, {Handberg}, {Hekker}, {Karoff}, {Kawaler}, {Lund}, {Lundkvist},
  {Metcalfe}, {Miglio}, {Silva Aguirre}, {Stello}, {White}, {Boss}, {Devore},
  {Gould}, {Prsa}, {Agol}, {Barclay}, {Coughlin}, {Brugamyer}, {Mullally},
  {Quintana}, {Still}, {Thompson}, {Morrison}, {Twicken}, {D{\'e}sert},
  {Carter}, {Crepp}, {H{\'e}brard}, {Santerne}, {Moutou}, {Sobeck}, {Hudgins},
  {Haas}, {Robertson}, {Lillo-Box}, \& {Barrado}}]{2014ApJS..210...20M}
{Marcy}, G.~W., {Isaacson}, H., {Howard}, A.~W., et~al. 2014, \apjs, 210, 20

\bibitem[{{Matsumura} {et~al.}(2010){Matsumura}, {Thommes}, {Chatterjee}, \&
  {Rasio}}]{2010ApJ...714..194M}
{Matsumura}, S., {Thommes}, E.~W., {Chatterjee}, S., \& {Rasio}, F.~A. 2010,
  \apj, 714, 194

\bibitem[{{Mazeh} {et~al.}(2013){Mazeh}, {Nachmani}, {Holczer}, {Fabrycky},
  {Ford}, {Sanchis-Ojeda}, {Sokol}, {Rowe}, {Zucker}, {Agol}, {Carter},
  {Lissauer}, {Quintana}, {Ragozzine}, {Steffen}, \&
  {Welsh}}]{2013ApJS..208...16M}
{Mazeh}, T., {Nachmani}, G., {Holczer}, T., et~al. 2013, \apjs, 208, 16

\bibitem[{{Minton} \& {Levison}(2014)}]{2014Icar..232..118M}
{Minton}, D.~A. \& {Levison}, H.~F. 2014, \icarus, 232, 118

\bibitem[{{Moore} {et~al.}(2013){Moore}, {Hasan}, \&
  {Quillen}}]{2013MNRAS.432.1196M}
{Moore}, A., {Hasan}, I., \& {Quillen}, A.~C. 2013, \mnras, 432, 1196

\bibitem[{{Murray} \& {Dermott}(1999)}]{1999ssd..book.....M}
{Murray}, C.~D. \& {Dermott}, S.~F. 1999, {Solar system dynamics}

\bibitem[{{Nagasawa} \& {Ida}(2011)}]{2011ApJ...742...72N}
{Nagasawa}, M. \& {Ida}, S. 2011, \apj, 742, 72

\bibitem[{{Ormel} {et~al.}(2012){Ormel}, {Ida}, \&
  {Tanaka}}]{2012ApJ...758...80O}
{Ormel}, C.~W., {Ida}, S., \& {Tanaka}, H. 2012, \apj, 758, 80

\bibitem[{{Petrovich} {et~al.}(2013){Petrovich}, {Malhotra}, \&
  {Tremaine}}]{2013ApJ...770...24P}
{Petrovich}, C., {Malhotra}, R., \& {Tremaine}, S. 2013, \apj, 770, 24

\bibitem[{{Rasio} \& {Ford}(1996)}]{1996Sci...274..954R}
{Rasio}, F.~A. \& {Ford}, E.~B. 1996, Science, 274, 954

\bibitem[{{Rein}(2012)}]{2012MNRAS.427L..21R}
{Rein}, H. 2012, \mnras, 427, L21

\bibitem[{{Rogers}(2014)}]{2014arXiv1407.4457R}
{Rogers}, L.~A. 2014, arXiv:1407.4457

\bibitem[{{Rowe} {et~al.}(2014){Rowe}, {Bryson}, {Marcy}, {Lissauer},
  {Jontof-Hutter}, {Mullally}, {Gilliland}, {Issacson}, {Ford}, {Howell},
  {Borucki}, {Haas}, {Huber}, {Steffen}, {Thompson}, {Quintana}, {Barclay},
  {Still}, {Fortney}, {Gautier}, {Hunter}, {Caldwell}, {Devore}, {Cochran},
  {Jenkins}, {Agol}, {Carter}, \& {Geary}}]{2014arXiv1402.6534R}
{Rowe}, J.~F., {Bryson}, S.~T., {Marcy}, G.~W., et~al. 2014, arXiv:1402.6534

\bibitem[{{Steffen} {et~al.}(2013){Steffen}, {Fabrycky}, {Agol}, {Ford},
  {Morehead}, {Cochran}, {Lissauer}, {Adams}, {Borucki}, {Bryson}, {Caldwell},
  {Dupree}, {Jenkins}, {Robertson}, {Rowe}, {Seader}, {Thompson}, \&
  {Twicken}}]{2013MNRAS.428.1077S}
{Steffen}, J.~H., {Fabrycky}, D.~C., {Agol}, E., et~al. 2013,
  \mnras, 428, 1077

\bibitem[{{Steffen} {et~al.}(2012){Steffen}, {Ford}, {Rowe}, {Fabrycky},
  {Holman}, {Welsh}, {Batalha}, {Borucki}, {Bryson}, {Caldwell}, {Ciardi},
  {Jenkins}, {Kjeldsen}, {Koch}, {Pr{\v s}a}, {Sanderfer}, {Seader}, \&
  {Twicken}}]{2012ApJ...756..186S}
{Steffen}, J.~H., {Ford}, E.~B., {Rowe}, J.~F., et~al.
  2012, \apj, 756, 186

\bibitem[{{Steffen} \& {Hwang}(2014)}]{2014arXiv1409.3320S}
{Steffen}, J.~H. \& {Hwang}, J.~A. 2014, arXiv:1409.3320

\bibitem[{{Weiss} \& {Marcy}(2014)}]{2014ApJ...783L...6W}
{Weiss}, L.~M. \& {Marcy}, G.~W. 2014, \apjl, 783, L6

\bibitem[{{Xie}(2014)}]{2014ApJS..210...25X}
{Xie}, J.-W. 2014, \apjs, 210, 25

\end{thebibliography}

\end{document}